\documentclass[a4paper,fleqn,usenatbib]{mnras}   

\usepackage{newtxtext,newtxmath}

\usepackage[T1]{fontenc}
\usepackage{ae,aecompl}

\usepackage{graphicx}	
\usepackage{amsmath}	
\usepackage{ulem}

\newcommand{\be}{\begin{equation}}
\newcommand{\ee}{\end{equation}}
\newcommand{\ba}{\begin{eqnarray}}
\newcommand{\ea}{\end{eqnarray}}
\newcommand{\hMpc}{{\ifmmode{h^{-1}\,{\rm Mpc}}
\else{$h^{-1}$Mpc}\fi}}
\newcommand{\Mpch}{h\,{\rm Mpc}^{-1}}
\newcommand{\bk}{\boldsymbol{k}}

\newcommand{\bx}{\boldsymbol{x}}

\newcommand{\nhal}{n_{\rm h}}

\title[The halo bispectrum]{The halo bispectrum as a sensitive probe of massive neutrinos and baryon physics} 

\author[V. Yankelevich et al.]{Victoria Yankelevich$^{1}$\thanks{E-mail: yankelevich.victoria@gmail.com},  Ian G. McCarthy$^{1}$\thanks{E-mail: I.G.McCarthy@ljmu.ac.uk},  
Juliana Kwan$^{1}$, Sam G. Stafford$^{1}$, Jia Liu$^{2}$ 
\\
$^{1}$Astrophysics Research Institute, Liverpool John Moores University, Liverpool, L3 5RF, UK\\
$^{2}$Kavli IPMU (WPI), UTIAS, The University of Tokyo, Kashiwa, Chiba 277-8583, Japan
}

\date{Accepted XXX. Received YYY; in original form ZZZ}

\pubyear{2018}

\begin{document}
\label{firstpage}
\pagerange{\pageref{firstpage}--\pageref{lastpage}}
\maketitle

\begin{abstract}
The power spectrum has been a workhorse for cosmological studies of large-scale structure.  However, the present-day matter distribution is highly non-Gaussian and significant cosmological information is also contained in higher-order correlation functions.  Meanwhile, baryon physics (particularly AGN feedback) has previously been shown to strongly affect the two-point statistics but there has been limited exploration of its effects on higher-order functions to date.  Here we use the \texttt{BAHAMAS} suite of cosmological hydrodynamical simulations to explore the effects of baryon physics and massive neutrinos on the halo bispectrum.  In contrast to matter clustering which is suppressed by baryon physics, we find that the halo clustering is typically enhanced.  The strength of the effect and the scale over which it extends depends on how haloes are selected.  On small scales ($k \ga 1$ $h$ Mpc$^{-1}$, dominated by satellites of groups/clusters), we find that the bispectrum is highly sensitive to the efficiency of star formation and feedback, making it an excellent testing ground for galaxy formation models.  We show that the effects of feedback and the effects of massive neutrinos are largely separable (independent of each other) and that massive neutrinos strongly suppress the halo bispectrum on virtually all scales up to the free-streaming length (apart from the smallest scales, where baryon physics dominates).  The strong sensitivity of the bispectrum to neutrinos on the largest scales and galaxy formation physics on the smallest scales bodes well for upcoming precision measurements from the next generation of wide-field surveys. 
\end{abstract}




\section{Introduction}

Providing a quantitative explanation for the formation and evolution of large-scale structure (LSS) in the Universe is one of the key tests of modern cosmological theories. The standard model of cosmology, the so-called $\Lambda$CDM model, has been remarkably successful in reproducing a wide range of observations, including the observed properties of the cosmic microwave background (e.g. \citealt{Planck2018}) and low-redshift probes such as baryon acoustic oscillations (e.g., \citealt{Eisensteien2005}) and redshift-space distortions (see e.g. \citealt{Alam2017}). While $\Lambda$CDM has been shown to reproduce most current LSS observations well, the physical nature of both dark matter and dark energy remain elusive.  Furthermore, there have been a number of recent mild tensions reported in the best-fit parameter values for certain cosmological parameters, including the Hubble constant (see \citealt{Verde2019} for a review) and the LSS parameter $S_8 \equiv \sigma_8 \sqrt{\Omega_m/0.3}$ (where $\sigma_8$ is defined as the amplitude of the (linear) power spectrum on the scale of $8 \, \hMpc$ and $\Omega_m$ is the present-day matter density parameter) as derived from measurements of cosmic shear, galaxy clustering, and the abundance of massive galaxy clusters (see, e.g., discussion in \citealt{McCarthy2018}).  Whether these tensions are hinting at the presence of (unaccounted for) systematic errors in some of the cosmological analyses, that there is new physics beyond the standard model, or that they represent moderately large statistical fluctuations, is presently unclear.

The drive to test $\Lambda$CDM further, and to look for new physics beyond the standard model, requires that we make increasingly precise measurements of `tried and tested' LSS statistics (e.g., 2-point clustering of galaxy clustering, weak lensing, etc.), but also that we devise new tests of LSS which are capable of probing different aspects of the matter distribution, which in turn could hopefully break important degeneracies between the fitted cosmological parameters.  The halo bispectrum (or 3-point function of haloes/galaxies), which we focus on in the present study, is one such example of a LSS test that is becoming increasingly observationally feasible (e.g., \citealt{PeeblesGroth1975, Gaztanaga1994, Scoccimarro+2001, Verde+2002, Gaztanaga+2005, PanSzapudi2005, Kulkarni+2007, McBride+2011,GilMartin2015a,GilMartin2015b,GilMartin2017,Slepian+2017a, Slepian+2017b,PearsonSamushia2018, Gualdi+2018, Gualdi+2019, Veropalumbo2021}) and for which previous theoretical studies have shown contains a significant source of cosmological information beyond what may be obtained from standard 2-point clustering tests alone \citep[e.g][]{Song15, Yankelevich, ChudaykinIvanov2019, Barreira2020, Gualdi+2020,MoradinezhadEtal2020, Hahn2020, HeinrichDore2020,   EggemeierEtal2021, Hahn2021, Nishant+2021}.

As in the case of most 2-point statistics, much of the observational signal of the bispectrum is expected to come from quasi-linear and non-linear scales.  Thus, `beyond linear' methods for evaluating the growth of structure, such as perturbation theory and full cosmological simulations, are generally required to predict the bispectrum. Examples of previous bispectrum work based on perturbation theory and cosmological simulations include \citet{Bernardeau+2002, Bernardeau+2008, Bernardeau+2012, CrocceScoccimarro2006, Pietroni2008, Taruya+2012, AnguloForeman+2015, BaldaufEFF+2015, Lazanu+2016,   Alkhanishvili+2021, SteeleBaldauf2021}. 

In addition to the extra complexity of following the gravitational evolution of matter when density fluctuations become large, there is also the challenge of accurately accounting for non-gravitational physics that comes into play.  Specifically, feedback processes associated with the formation of stars and the growth of black holes are well-established, both empirically and via theoretical models and full cosmological hydrodynamical (hereafter hydro) simulations, to have a strong effect on the overall matter distribution on scales of up to a few tens of Mpc (e.g., \citealt{vanDaalen11, Schneider2015, Mummery17, Chisari19, vanDaalen2020}). Previous studies that focused on 2-point statistics have shown that, unless these effects are properly modelled, they will result in biases in the derived cosmological parameters (e.g., \citealt{Semboloni2011}).

Given that galaxy formation physics has been shown to significantly affect classical 2-point statistics, it is reasonable to expect that it should also affect the 3-point correlation function/bispectrum. But how large are the effects? Is the bispectrum more or less sensitive to these processes than previously studied LSS quantities? And how do we mitigate against these effects (if the goal is cosmological constraints) or exploit them (if the goal is to better understand galaxy formation)?  These are the questions we focus on in the present study.

Note that, thus far, there has been remarkably little attention devoted to the effects of baryon physics on the halo bispectrum using cosmological simulations.  This is plausibly explained by the fact that very large cosmological volumes (box sizes of at least several hundred Mpc on a side) are required to reliably measure the bispectrum.  Combined with the constraint that the simulations must also use full hydrodynamics and a model for galaxy formation in order to self-consistently incorporate the impact of baryons, there are very few simulations that presently meet these criteria.  In fact, we are aware of only two suites of hydro simulations of sufficient volume for this purpose: the \texttt{BAHAMAS} simulations \citep{Mccarthy2017,McCarthy2018} and the Magneticum simulations \citep{Dolag2016,Bocquet2016}.  However, to date neither have been used to examine the halo/galaxy bispectrum. 
\citet{Semboloni+2013} presented a study of baryonic effects on the \textit{matter} bispectrum using the \texttt{OWLS} simulations  \citep{Schaye2010}.  Their study also showed how the combination of the power spectrum and bispectrum can be used to break degeneracies between baryonic and cosmological effects. 
\citet{Foreman+20}, have used \texttt{BAHAMAS} to also explore the effects of baryons on the matter bispectrum, finding that baryon physics does significantly affect the matter bispectrum and that the effects differ in magnitude and scale dependence compared to their effects on the matter power spectrum.  \citet{Arico2021} used the measurements of \citet{Foreman+20} to show that their `baryonification' method for altering the outputs of collisionless (DM-only) simulations can be used to simultaneously model the effects of baryons on the matter power spectrum and the reduced matter bispectrum.  We note here that, while the matter bispectrum is related to the halo bispectrum (and that the former is considerably easier to measure in simulations, given a large number of tracer particles), they are clearly distinct quantities, as haloes/galaxies are biased tracers of the matter distribution.  Indeed, we will show later that the impact of baryons on halo clustering is qualitatively different (opposite sign) to that on the matter distribution.  Note that, observationally, the halo/galaxy bispectrum may be measured from galaxy redshift surveys, whereas the matter bispectrum can be inferred from a combination of galaxy and weak lensing (cosmic shear) surveys (e.g., \citealt{Fu2014}).

In this paper, we use the \texttt{BAHAMAS} suite of cosmological hydro simulations to explore the impact of baryon physics (particularly active galactic nucleus (AGN) feedback) on the halo bispectrum.  We also examine the role of massive neutrinos, noting that previous theoretical studies have highlighted the sensitivity of the bispectrum to this component (e.g., \citealt{Hahn2020,Hahn2021,Chen2021}), and the potential degeneracy between baryons and neutrinos.

The paper is structured as follows.  In Section \ref{BAHAMAS} we describe the cosmological simulations used in this study.  In Section \ref{bisp} we describe our methods for measuring the halo power spectrum and bispectrum and, for the latter, discuss the dependence on triangular configuration.  In Section \ref{results} we present our main results, showing the dependence of the bispectrum on baryon physics, halo mass, and neutrino mass, and we explore the degeneracy between baryons and neutrinos.  In Section \ref{conclusions} we summarise and discuss our findings.

\section{BAHAMAS}
\label{BAHAMAS} 

In the present study, we use the \texttt{BAHAMAS}\footnote{\url{https://www.astro.ljmu.ac.uk/~igm/BAHAMAS/}} cosmological simulations \citep{Mccarthy2017,McCarthy2018} to explore the impact of baryons and neutrinos on the halo bispectrum. \texttt{BAHAMAS} is a suite of large-volume hydro simulations that simultaneously explores variations in baryonic physics and cosmology, including extensions to $\Lambda$CDM such as massive neutrinos \citep{Mummery17}, dynamical dark energy \citep{Pfeifer2020}, and a running scalar spectral index \citep{Stafford2020a}.  Here we use 7 runs from \citet{McCarthy2018} which explore variations in AGN feedback and the summed mass of the neutrinos in the context of a WMAP 9-year cosmology \citep{WMAP9}.  Table~\ref{tab:sims} presents the cosmological parameters of \texttt{BAHAMAS} simulations used in the current work.

Each of the hydro runs used in this study has a box size of $400$ $\hMpc$ and includes $2 \times 1024^3$ particles, equally split between baryons and dark matter, which have a mass ratio equal to $\Omega_{\rm b}/\Omega_{\rm cdm}$.  (As described below, the neutrinos are not followed using particles.)  In addition, for each hydro run, there is a corresponding collisionless simulation which, as described in \citet{vanDaalen2020}, is represented by two collisionless fluids (one for the baryons and one for the dark matter).

The Boltzmann code {\small CAMB}\footnote{\url{http://camb.info/}} \citep{CAMB1, CAMB2} was used to compute the transfer functions and a modified version of {\small N-GenIC} was used to create the initial conditions at a starting redshift of $z=127$.  {\small N-GenIC} was modified to include second-order Lagrangian Perturbation Theory corrections and support for massive neutrinos\footnote{\url{https://github.com/sbird/S-GenIC}}.  Separate transfer functions, computed by {\small CAMB}, are used for each individual component (baryons, neutrinos, and CDM) when producing the initial conditions.  Note also that the same random phases are used for each of the simulations, such that comparisons between different runs are not subject to cosmic variance.

The simulations were carried out with a version of the Lagrangian TreePM-SPH code \textsc{gadget3} \citep[last described in][]{Springel2005}, which was modified to include subgrid physics as part of the \texttt{OWLS} project \citep{Schaye2010}.  The gravitational softening is fixed to $4~h^{-1}$ kpc (in physical coordinates below $z=3$ and in comoving coordinates at higher redshifts) and the SPH smoothing is done using the nearest 48 neighbours.  

To include the effects of massive neutrinos, \texttt{BAHAMAS} uses the semi-linear algorithm developed by \citet{Bird2013}.  The algorithm evaluates neutrino perturbations on the fly at every time step using a linear perturbation integrator, which is sourced from the full non-linear baryons+CDM potential and is added to the total gravitational force.  The dynamical responses of the neutrinos to the baryons+CDM and of the baryons+CDM to the neutrinos are mutually and self-consistently included. 

The hydro simulations include subgrid prescriptions for metal-dependent radiative cooling, star formation and stellar evolution (including chemical enrichment from Type II and Ia supernovae and Asymptotic Giant Branch stars).  The simulations include stellar feedback and prescriptions for supermassive black hole growth and AGN feedback.  See \citet{Schaye2010} and references therein for a detailed description of the subgrid implementations.

As described in \citet{Mccarthy2017}, the parameters controlling the efficiencies of the stellar and AGN feedback were adjusted so that the simulations reproduce the present-day galaxy stellar mass function (GSMF) for $M_* > 10^{10}$ M$_\odot$ and the amplitude of the gas mass fraction$-$halo mass relation of groups and clusters, as inferred from high-resolution X-ray observations.  These quantities were selected to ensure that the collapsed structures in the simulations have the correct baryon content in a global sense, which \citet{vanDaalen2020} have shown is critical for capturing the impact of baryons on the matter power spectrum. 
We point out that the parameters governing the feedback efficiencies were {\it not} recalibrated when varying the cosmological parameters away from the fiducial WMAP~9-yr cosmology (with massless neutrinos).  But as shown in \citet{McCarthy2018}, the internal properties of collapsed structures (stellar masses, gas masses, etc.) are, to first order, insensitive to the variations in cosmology.  Nevertheless, we explore the potential degeneracy between baryon physics and neutrino free-streaming in Section~\ref{sec:neutrinos}.

\begin{table*} 
\caption{\label{tab:sims} Cosmological parameter values for the \texttt{BAHAMAS} simulations used here. For all the models a WMAP~9-based cosmology is adopted. 
The columns are: (1) model name; (2) summed mass of the 3 active neutrino species (we adopt a normal hierarchy for the individual masses); (3) subgrid AGN heating temperature; (4) Hubble's constant; (5) present-day baryon density; (6) present-day dark matter density; (7) present-day neutrino density, computed as $\Omega_\nu = M_\nu / (93.14 \ {\rm eV} \ h^2)$; (8) spectral index of the initial power spectrum; (9) amplitude of the initial matter power spectrum at a {\small CAMB} pivot $k$ of $2\times10^{-3}$ Mpc$^{-1}$; (10) present-day (linearly-evolved) amplitude of the matter power spectrum on a scale of 8 $\hMpc$ (note that we use $A_s$ rather than $\sigma_8$ to compute the power spectrum used for the initial conditions, thus the ICs are `CMB normalised').  In addition to the cosmological parameters, we also list the following simulation parameters: (11) dark matter particle mass; (12) initial baryon particle mass. }
\begin{tabular}{*{12}{c}}                                                                 \hline
(1)        & (2)        & (3)               & (4)                & (5)              & (6)      & (7)              &  (8)          &   (9)                       & (10) & (11)    & (12)                   \\
Model & $M_{\nu}$ & $\Delta T_{\rm {heat}}$  & $H_0$      & $\Omega_b $  & $\Omega_{\rm cdm} $ &  $\Omega_\nu$     & $n_{\rm s}$    & $A_s$            &  $\sigma_8$   &  $M_{\rm DM}$                & $M_{\rm bar,init}$           \\
& (eV)    & (K)   & (km/s/Mpc) &                  &                    &                  &          & ($10^{-9}$)      &               &  [$10^9 h^{-1} {\rm M_{\odot}}$]  & [$10^8 h^{-1} {\rm M_{\odot}}$]   \\
\hline
fiducial & 0.0  & $10^{7.8}$   & 70.00 & 0.0463 & 0.2330 & 0.0    & 0.9720 & 2.392 & 0.8211 & 3.85 & 7.66 \\
low AGN & 0.0  & $10^{7.6}$   & 70.00 & 0.0463 & 0.2330 & 0.0    & 0.9720 & 2.392 & 0.8211 & 3.85 & 7.66 \\
high AGN & 0.0  & $10^{8.0}$   & 70.00 & 0.0463 & 0.2330 & 0.0    & 0.9720 & 2.392 & 0.8211 & 3.85 & 7.66 \\
0.06 & 0.06    & $10^{7.8}$  & 70.00 & 0.0463 & 0.2317 & 0.0013 & 0.9720 & 2.392 & 0.8069 & 3.83 & 7.66 \\
0.12 & 0.12   & $10^{7.8}$   & 70.00 & 0.0463 & 0.2304 & 0.0026 & 0.9720 & 2.392 & 0.7924 & 3.81 & 7.66 \\
0.24 & 0.24   & $10^{7.8}$   & 70.00 & 0.0463 & 0.2277 & 0.0053 & 0.9720 & 2.392 & 0.7600 & 3.77 & 7.66 \\
0.48 & 0.48   & $10^{7.8}$   & 70.00 & 0.0463 & 0.2225 & 0.0105 & 0.9720 & 2.392 & 0.7001 & 3.68 & 7.66 \\
\hline
\end{tabular}
\end{table*}

\section{Bispectrum estimation}
\label{bisp} 
\subsection{Power spectrum and bispectrum definitions}

The power spectrum and the bispectrum are the Fourier transforms of two- and three-point correlation functions, respectively, of the dimensionless overdensity $\delta(\bx)$ parameter, which is defined as:
\be
\delta( \bx) = \frac{\rho(\bx)}{\bar \rho}-1\; , 
\ee
where $\rho(\bx)$ is a continuous random field which gives the local density per unit comoving volume in the expanding Universe and $\bar \rho$ is the mean density with $\bar{\rho}=\langle \rho(\bx)\rangle$ (the brackets $\left\langle \cdots \right\rangle$ denotes the averages taken over an ideal ensemble of realisations). 

We define the power spectrum and the bispectrum, respectively, as:
\be
\left\langle\delta( \bk) \, \delta( \bk')\right\rangle = (2\pi)^3 \,P( \bk) \, \delta_{\rm D}( \bk +  \bk')\;,
\label{def_ps}
\ee
\be
\left\langle\delta( \bk_1)\,\delta( \bk_2) \, \delta( \bk_3)\right\rangle =(2\pi)^3\, B( \bk_1, \bk_2, \bk_3)\, \delta_{\rm D}( \bk_{123})\;,
\label{def_bisp}
\ee
where $\delta_{\rm D}(\bk)$ is the three dimensional Dirac delta function 
with $ \bk_{123}=\bk_1 +  \bk_2+  \bk_3$, which implies that the bispectrum is defined only for closed triangles of wavevectors.

The halo power spectrum and the halo bispectrum measured from numerical simulation or observational data can be significantly affected by shot noise. 
Assuming the distribution of haloes derives from Poisson sampling, the measured spectra (which is a sum of the true spectra and shot noise terms) can be defined as \citep[e.g.][and references therein]{Oddo+},:
\be
\tilde{P}(\bk)=P(\bk)+\frac{1}{\nhal}\;,
\label{eq:snP}
\ee
\begin{multline}
 \tilde{B}(\bk_1,\bk_2,\bk_3)=B(\bk_1,\bk_2,\bk_3)\\
+\left[P(\bk_1)+P(\bk_2)+P(\bk_3)\right]\frac{1}{\nhal}+\frac{1}{\nhal^2}\;,
\label{eq:snB}
\end{multline}
where $\nhal$ is the halo number density \citep[e.g.][]{Matarrese+97}.  Note that the assumption of a Poisson distribution for the bispectrum shot noise term is accurate to first-order only, since the clustering of haloes is not strictly Poissonian. We discuss in detail the role of Poisson and non-Poisson shot noise in Section \ref{sec:shot_noise}.

\subsection{Halo power spectrum and bispectrum measurements}

In order to measure halo power spectra and bispectra, we modify the publicly-available \textsc{bskit}\footnote{\url{https://github.com/sjforeman/bskit}} code of \citet{Foreman+20}.   The original code measures the matter power spectrum and matter bispectrum from the \texttt{BAHAMAS}, \texttt{IllustrisTNG}\footnote{\url{https://www.tng-project.org}}, \texttt{Illustris}\footnote{\url{http://www.illustris-project.org}} and \texttt{EAGLE}\footnote{\url{http://icc.dur.ac.uk/Eagle/}} simulations.
 \textsc{bskit} is built upon the \textsc{nbodykit} \citep{nbodykit} simulation analysis package, and uses the standard FFT-based bispectrum estimator \citep{Scoccimarro2000, Sefusatti+2016, Tomlinson+2019FFT}.
By default, the \texttt{BAHAMAS} module of \textsc{bskit} reads in the positions and masses of particles in order to compute the matter power spectrum and bispectrum. 
We modify it for this study to read in the positions (defined here as the centre of potential) and masses of haloes as opposed to particles.  When computing the halo power spectra and bispectra with \textsc{bskit} by default we weight each halo equally.  In Appendix \ref{App:weight} we explore the effects of weighting haloes by their mass and show that such a weighting scheme tends to reduce the impact of baryons on the bispectrum, as higher mass haloes, which are less affected by ejective AGN feedback, contribute more to the computed spectra.

In this work, we analyse all self-gravitating substructures, often referred to as `subhaloes', regardless of whether they are central or satellite subhaloes. Such a selection is closer to a galaxy-based selection than selecting central subhaloes only since all subhaloes of sufficient mass are expected to host galaxies. 
Without risk of confusion, going forward we will collectively refer to such self-gravitating substructures as just `haloes'.  For \texttt{BAHAMAS}, a standard friends-of-friends (FoF) algorithm \citep{FoF-Davis-Peebles1983} was run to identify FoF groups and the \texttt{SUBFIND} \citep{Springel+2001-subfind, Dolag+2009-subfind} algorithm was then used to identify all self-gravitating substructures (haloes). 

Note that, because the simulations use the same random phases in the initial conditions, it is possible to identify the same haloes in both the hydro run and its corresponding collisionless run.  This allows us to isolate the impact of baryonic and cosmological effects in a straightforward way.  We refer to the case where we analyse a common set of haloes between multiple simulations as a `matched' sample.  An `unmatched' sample refers to the case where we simply select haloes based on some criteria (e.g., total mass) from each simulation without regard for whether they correspond to the same haloes\footnote{In general they will not be the same, since feedback and cosmological effects alter the masses of haloes.}.  Observationally, it is often the case that haloes/galaxies are selected by, for example, their total masses, thus it is also useful to analyse the results for the unmatched case. 

To construct matched samples of haloes, we use the unique dark matter particle IDs, using the 50 most bound particles in each halo of a reference simulation (e.g., the collisionless run).  Whichever halo in the target simulation (e.g., the hydro simulation) contains the largest fraction of these dark matter particles from the reference simulation is considered that haloes match. Matches are performed bijectively, however, and only haloes that are matched both backwards and forwards are retained in the final matched sample.

The size of the simulation box and its mass resolution limits the range of scales over which the power spectra and bispectra can be measured.  $k_{\rm f}=2\pi/L_{\rm{box}}=0.0157 \Mpch$ corresponds to the so-called the fundamental $k$-mode of the \texttt{BAHAMAS} simulations, and is the longest wavelength mode that can be measured given the finite box size. In terms of the binning strategy between these limits, the choice of bin width, $\Delta k $, represents a trade-off between the accuracy of results and the computational cost.  Note that the computational cost comes in two forms, which is the evaluation of the spectra from a grid (with finer bins requiring more evaluations) and the construction of the grid itself, which can be the most costly aspect depending on the required resolution.
It has been shown before that the bispectrum is sensitive to the size of $\Delta k $ in various ways \citep[e.g.][]{Yankelevich, Oddo+}. 
In many theoretical works, a binning of $\Delta k =k_{\rm f}$ is used. 
However, such narrow bins may not be the best choice for measuring the bispectrum from simulations, as the computational cost scales inversely with the width of the k bins used. 
In addition to this, narrow $k$-bins also leads to additional numerical noise. 
The common solution to this problem is to have narrow (wide) $k$-bin for small (large) $k$.
For example, \citet{Foreman+20} use $\Delta k =k_{\rm f}$ for $k<40k_{\rm f}$ and $\Delta k =6 k_{\rm f}$ for $40k_{\rm f}<k<k_{\rm max}$, where $k_{\rm max}$ is the maximum wave-vector for which the analysis is done. We adopt $k_{\rm max}=3 \Mpch$ throughout.  

Fortunately, the computational resources to measure the halo bispectrum are lower than for the matter bispectrum in the example above, since the construction of the grid is considerably cheaper for the former. 
Therefore, when computing the halo bispectrum we use the same bin size of $\Delta k =2k_{\rm f}$ over the entire range of $k$-scales.  However, note that when plotting power spectrum and bispectrum ratios in the figures described later, we apply some smoothing to the curves to reduce the impact of noise.  Specifically, we rebin the spectra by a factor 3 and apply a Savitzky-Golay filter.

\begin{figure*}
	\includegraphics[width=14cm]{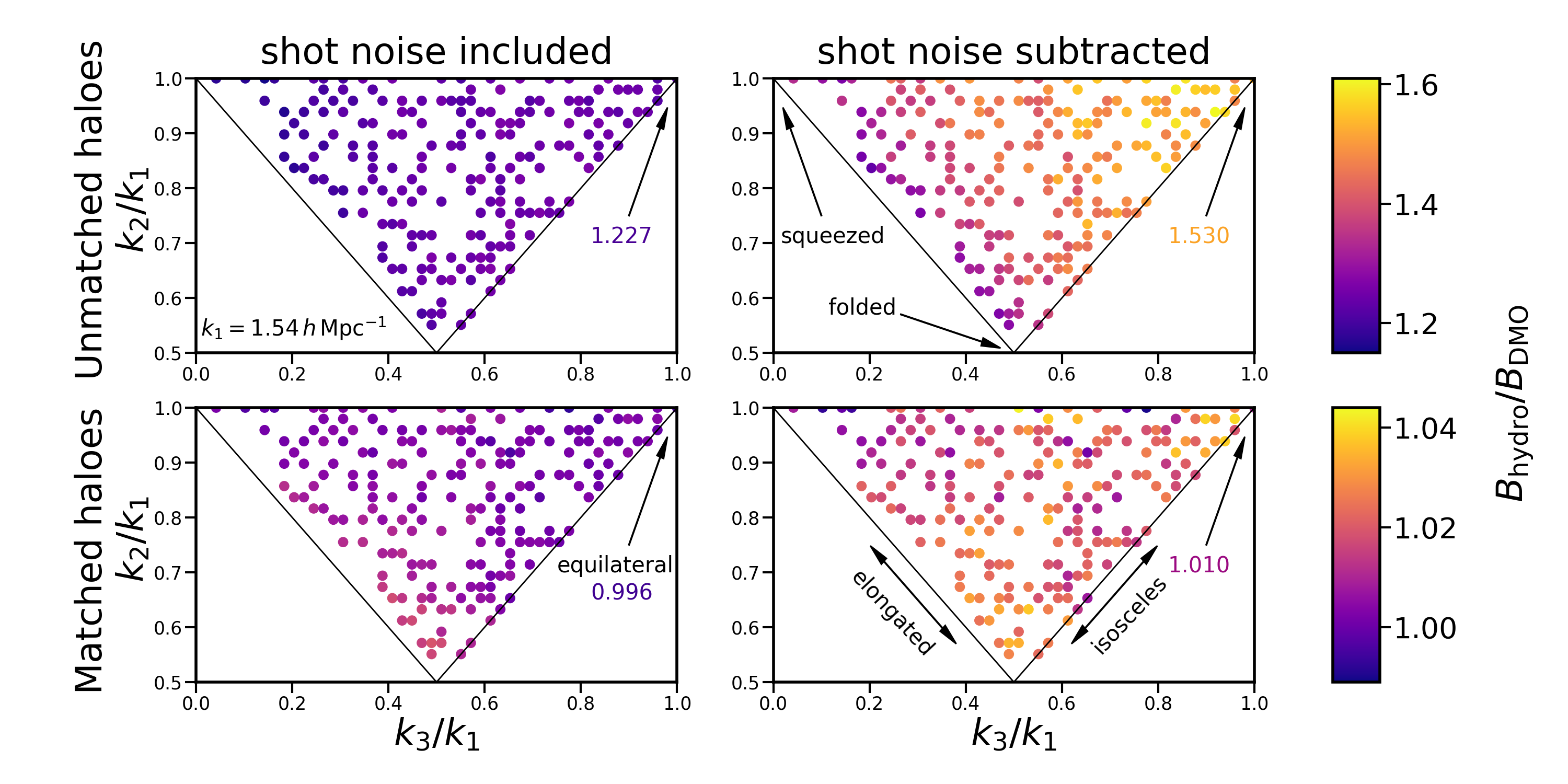}
    \caption{The colour-coded value of the ratio of the bispectrum, $B_{\rm{hydro}} / B_{\rm {DMO}}$, measured from the fiducial hydro and dark matter-only \texttt{BAHAMAS} simulations for different triangular configuration for the fixed value of $k_1=1.54\, \Mpch$. The ratio is presented for unmatched (top row) and matched (bottom raw) haloes with masses $M \geq 10^{11.5}h^{-1}$ M$_{\odot}$, with and without shot noise subtraction (right and left columns, respectively). The colour-coded numbers inside each subplot show the value of the ratio for the equilateral triangular configuration. The arrows indicate the positions of various triangular configurations. 
}
    \label{fig:diff_tri_config}
\end{figure*}

\subsection{Choice of triangular configuration}
\label{sec:Choice of triangular configuration}

A potentially important question is what triangular configuration $\bk_1,\bk_2,\bk_3$ should be adopted when computing the bispectrum.  Results in the literature are often presented for the equilateral case only. However, there is no particularly strong motivation behind this choice, except perhaps for convenience.  We present here a brief analysis of the dependence of (the baryons effects on) the halo bispectrum on the choice of triangular configuration.

In Fig.~\ref{fig:diff_tri_config} we show the dependence of the effects of baryonic physics on the halo bispectrum as a function of triangular configuration (note that we discuss the physical origin of these effects in Section \ref{results}). We do this by plotting the ratio of the bispectrum measured from the fiducial hydro simulation with respect to the corresponding collisionless (DMO) simulation for the case with massless neutrinos.

For this test, we calculate the bispectrum for all haloes with masses above $10^{11.5}h^{-1}$ M$_{\odot}$. 
This choice, which roughly corresponds to a stellar mass of $10^{10}$ M$_{\odot}$, is dictated primarily by the mass resolution of the simulations.  
For the unmatched case, this mass cut is applied to both the DMO and hydro simulations separately, and thus somewhat different populations of haloes are selected in the two runs.  For the matched case, however, we apply the mass cut to the DMO simulation and then select the corresponding matched haloes from the hydro simulation.  For the purposes of this comparison we fix $k_1$ to $1.54 \, \Mpch$, since
(i) there is an equilateral triangle with such side length; (ii) there is a sufficient number of independent triangles such that the bispectrum is well measured; and (iii) it falls approximately half way between the full range of scales that we consider in the rest of the paper ($0.0157 \, \Mpch$ to $3 \,  \Mpch$).  In Appendix \ref{App:tri_choice} we explore how the choice of triangular configuration impacts the results for different choices of $k_1$.

The dots in Fig.~\ref{fig:diff_tri_config} (and in Figs.~\ref{fig:diff_tri_config_low_k}-\ref{fig:diff_tri_config_high_k}) appear in an irregular way because the bispectrum measurements from the simulations do not contain all the possible triangular configurations for the chosen values of $k_1$. We adopt the same order of wave-vector length as in \textsc{bskit} which is $k_1\geq k_2 \geq k_3$. For theoretical calculations all the triangles which follow triangles' condition on the sides, in our case, $k_1 \leq k_2+k_3$ would exist and will fill all the ${k_2/k_1 - k_3/k_1}$ space. However, not all the configurations exist inside the simulation and therefore there are some gaps in the distribution.

The value of the $B_{\rm{hydro}} / B_{\rm {DMO}}$ in Fig.~\ref{fig:diff_tri_config} is colour-coded using two colour bars: one for the unmatched haloes (top row) and another for the matched haloes (bottom row) cases. The arrows pointing to the top right corner of each subplot indicate the ratio for equilateral triangle configuration. The right (left) column shows the results for the case where the shot noise has (not) been subtracted.

We can see from Fig.~\ref{fig:diff_tri_config} that the impact of the choice of configuration appears to depend on the selection of haloes adopted (unmatched vs.~matched) and on whether the bispectrum has been shot noise subtracted. For the shot noise-subtracted case with unmatched haloes (arguably the most useful comparison, since the spectra should be shot noise subtracted and the unmatched case is closer to an observational selection), the ratio of the bispectrum ranges from 1.2 to 1.6, with the equilateral case yielding close to the maximal ratio. The fact that the ratio depends on the choice of triangular configuration suggests that it may be possible to extract additional information about the impact of baryons by computing the bispectra for a range of different configurations. We plan to explore this possibility further in future work and proceed with the standard equilateral configuration for the remainder of this study, noting that there is some sensitivity to the results on the choice of configuration.

\subsection{Shot noise and resolution considerations}
\label{sec:shot_noise}

\begin{figure*} 
	\includegraphics[width=\columnwidth]{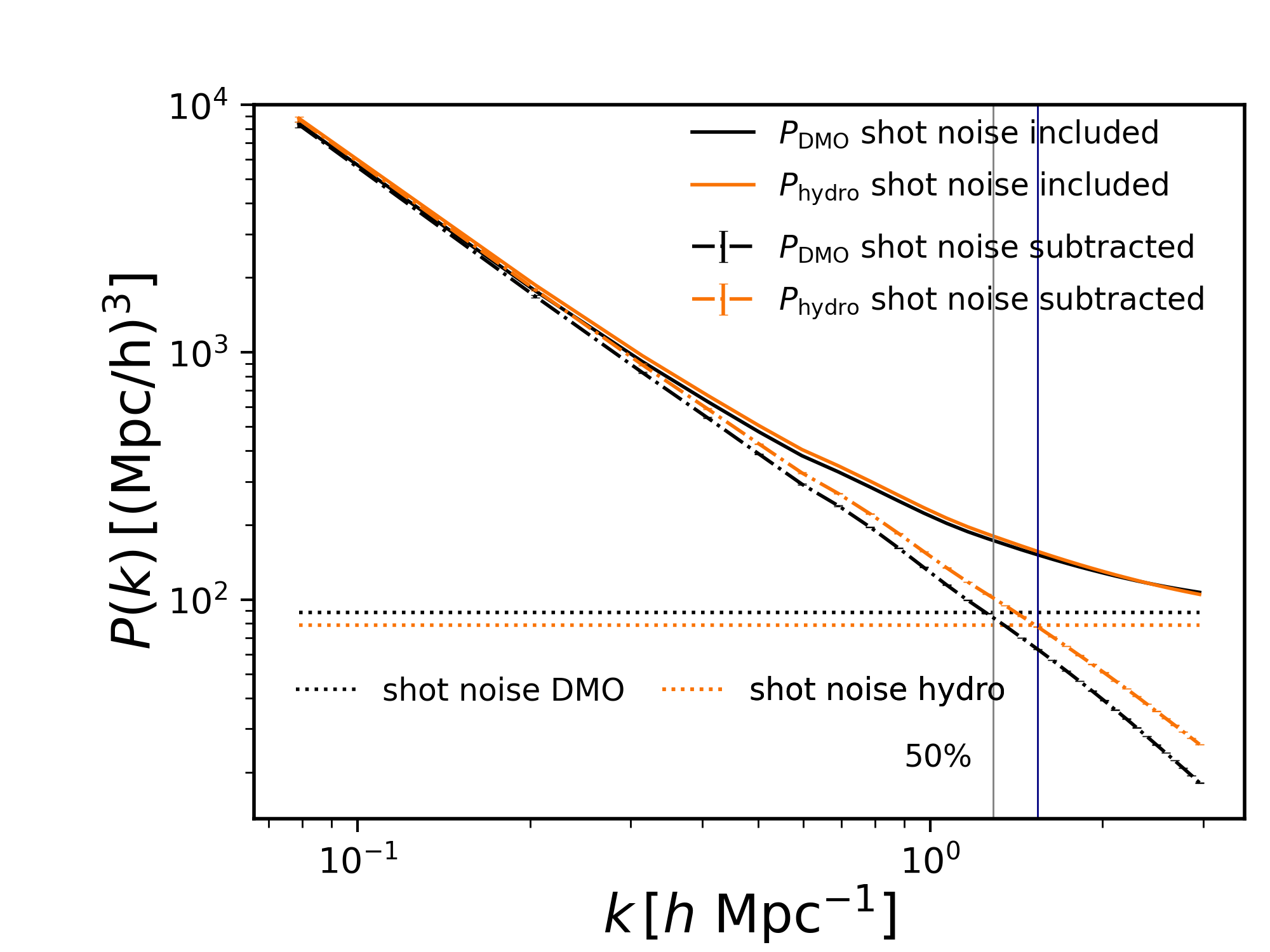}
    \includegraphics[width=\columnwidth]{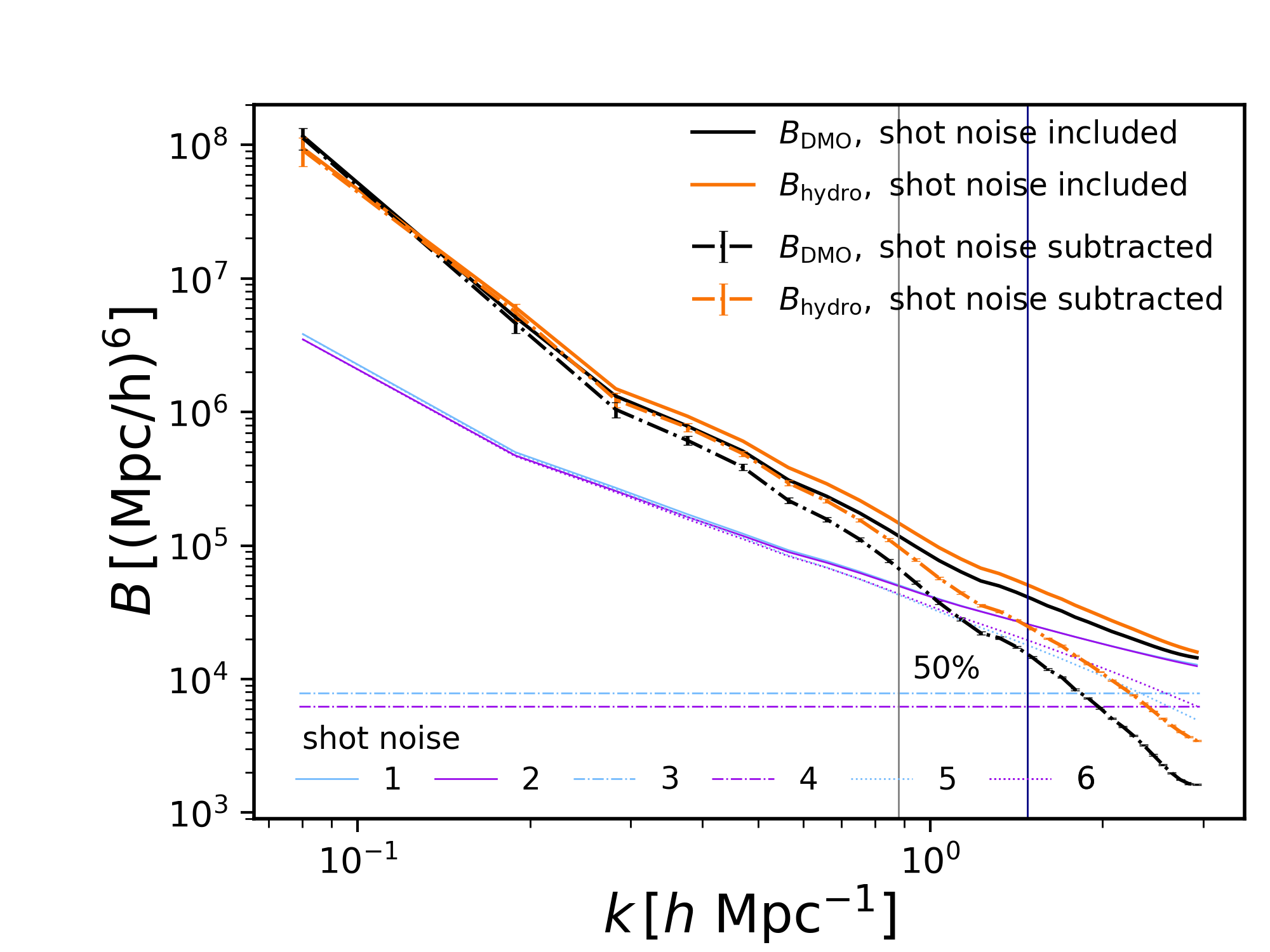}
  \caption{ The power spectrum and the bispectrum measured from the fiducial hydro (orange lines) and DMO (black lines) BAHAMAS simulations
with (dashdot lines with errorbars) and without (solid lines) shot noise subtraction assuming equilateral triangles. The haloes mass cut is the same as in our fiducial analysis (e.g., Fig.~\ref{fig:p_b_ratio_general}) for the ‘unmatched’ selection. The error bars indicate the theoretically estimated Gaussian errors.  In the left panel the dotted lines represent the Poisson shot noise for the power spectrum (the last term in equation \ref{eq:snP}).  In the right panel the shot noise is shown according to the legend: 1 -- full shot noise for DMO bispectrum (second and third terms in equation \ref{eq:snB}), 3 -- part of shot noise for DMO bispectrum (third term in equation \ref{eq:snB}), 5 -- part of shot noise for DMO bispectrum (second term in equation \ref{eq:snB}), all cyan colour. The purple lines 2,4,6 are the same as 1,3,5 but for the hydro bispectrum.  The two vertical lines in both panels indicate the 50\% shot noise contribution for DMO (grey colour) and hydro (navy colour) power spectrum and bispectrum. } 
  \label{fig:ps_bisp_general} 
\end{figure*} 

Here we present an analysis of the impact of shot noise on our results and its dependence on the adopted triangular configuration.  We also discuss the possible impact of non-Poisson shot noise as well as possible limitations due to finite mass and force resolution as well as finite volume effects.  In short, we conclude that resolution and box size limitations are unlikely to be important for our results and conclusions, while the contribution of shot noise to the halo power spectra and bispectra becomes important (i.e., comparable to the true clustering signal) on scales of $k \approx 1 \, \Mpch$.   Thus, a degree of caution is warranted when examining the absolute power beyond this scale, although we expect the \textit{relative} trends in $P(k)$ and $B(\bk_1,\bk_2,\bk_3)$ (e.g., as we increase feedback or the summed neutrino mass) to be robust.  Readers who are mainly interested in the impact of baryon physics and neutrino free-streaming on the halo power spectrum and bispectrum may wish to skip ahead to Section \ref{results}.

We first consider the impact of Poisson shot noise.  In Fig.~\ref{fig:ps_bisp_general} we show the absolute power spectrum and bispectrum (both DMO and hydro) with and without shot noise subtraction, which illustrates clearly the relative contribution of shot noise as a function of scale. In the case of the bispectrum, we show the two Poisson terms separately (see equation \ref{eq:snB} for details).  The plot also shows individual data points with a theoretical estimate of the errors in the Gaussian approximation for both power spectrum and bispectrum.  For the theoretical Gaussian errors we use a standard approach where the error is the square root of the covariance matrix. For details of how to calculate the covariance matrix see equations 2.34 and 2.39 in \citet{Yankelevich}.

Two vertical lines are added to Fig.~\ref{fig:ps_bisp_general} to indicate the 50\% shot noise contribution for both the power spectrum and bispectrum.  (The $k$ values at which Poisson shot noise contributes $1\%$ and $10\%$ are also presented in Table~\ref{tab:k_val}.) 
From this figure we conclude that the contribution from Poisson shot-noise becomes comparable to that of the true clustering signal at scales of $k \approx 1 \, \Mpch$.  Thus, a degree of caution is warranted when examining the absolute power beyond this scale, although we expect the \textit{relative} trends (e.g., as we increase feedback or the summed neutrino mass) to be robust.

\begin{figure*}
	\includegraphics[width=14cm]{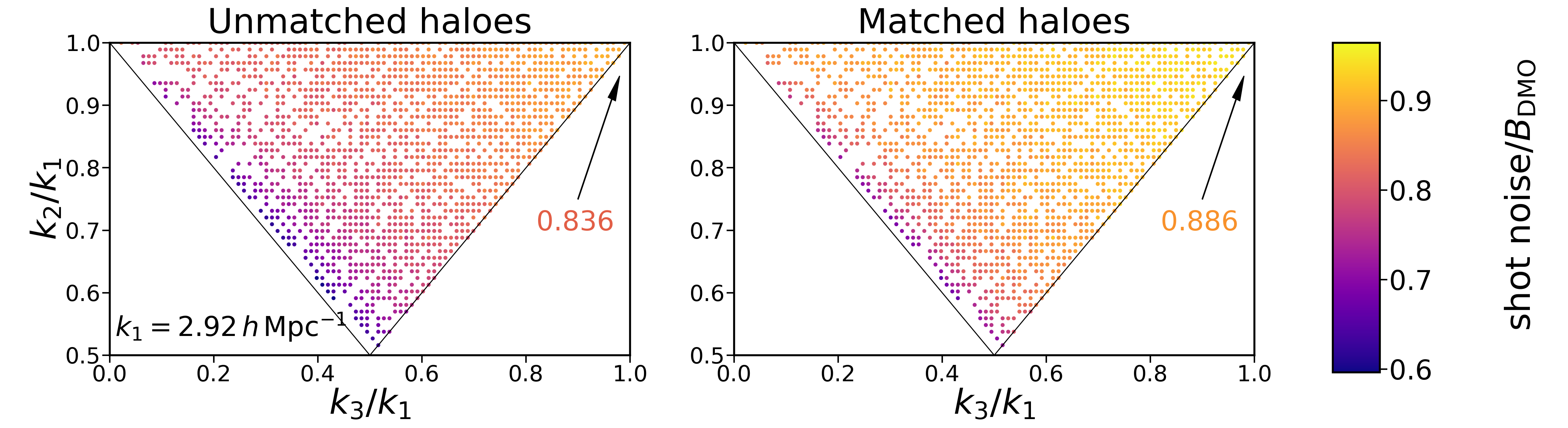}
    \caption{The relative contribution of the Poisson shot-noise to the bispectrum as a function of the triangular configuration. For consistency with Fig.~\ref{fig:diff_tri_config_high_k} we use the same $k_1=2.92\, \Mpch$.
}
    \label{fig:sn_bisp_tri}
\end{figure*}

Note that the shot noise contribution to the bispectrum depends on the adopted triangular configuration.  Fig.~\ref{fig:sn_bisp_tri} demonstrates that the shot noise ratio is smaller for the squeezed, elongated and folder triangles, and is generally larger for isosceles and equilateral configurations. The largest shot noise case corresponds to the scalene configuration.

\begin{table*} 
\caption{\label{tab:k_val} Values of wave-vector $k$ at which the shot noise contributes $1\%$, $10\%$ or $50\%$ to the absolute power spectrum or bispectrum, and the values of the shot noise (SN) itself. Note that the units of shot noise are $h^{-3}\rm{Mpc}^3$ for the power spectrum and $h^{-6}\rm{Mpc}^6$ for the bispectrum.   }
\begin{tabular}{*{7}{c}}                              \hline
 & $k, 1\%$ & SN & $k, 10\%$ & SN & $k, 50\%$ & SN  \\
 & $\Mpch$ &  & $\Mpch$ & & $\Mpch$ &\\
\hline
$P_{\rm{DMO}}$ & 0.063 & 88.71 & 0.346 & 88.71 & 1.288 & 88.71\\
$P_{\rm{hydro}}$ & 0.094 & 78.91& 0.377 &78.91 & 1.539 & 78.91\\ 
$B_{\rm{DMO}}$ & 0.064 & $2.24 \cdot 10^6$ & 0.064 & $2.24 \cdot 10^6$ & 0.879 & $5.06 \cdot 10^4$ \\
$B_{\rm{hydro}}$ & 0.036 & $3.74 \cdot 10^6$ & 0.064 & $2.09 \cdot 10^6$ & 1.476 & $2.58  \cdot 10^4$ \\
\hline
\end{tabular}
\end{table*}  

In our fiducial analysis in Section \ref{results}, we consider the Poissonian shot noise contribution only.  However, on small scales we expect the pure Poissonian assumption to break down since haloes can be strongly clustered (e.g., satellites in a galaxy group).  Previous work has shown that the non-Poisson contribution to the shot noise is generally subdominant to that of the Poissonian component.  For example, using the halo model, \citet{Ginzburg+2017} have calculated that the non-Poisson shot noise is approximately half that of the Poisson shot noise contribution at largest $k$ values (smallest scales) considered, a result which they confirmed using N-body simulations.  At smaller $k$ values (large scales), the contribution from non-Poisson shot noise is generally negligible.  Thus, we expect that, at worse, our estimate of the shot noise is biased low by 50\% and, therefore, our estimate of the true intrinsic bispectrum signal small scales ($k \approx 1 \, \Mpch$) may be biased high up to 25\%.  However, we do not expect our relative findings (i.e., ratios of bispectra) to be as adversely affected, since the absolute spectra will all be similarly affected.

We now turn our attention to resolution considerations.  Note that the force resolution limit (taken to be the Plummer equivalent gravitational softening of 4 kpc $h^{-1}$) of the simulations is $k \approx 1507 \, \Mpch$, which is well beyond the scales we analyse.  Thus, limitations due to force resolution are unimportant for our analyses.  However, the force resolution is not the only resolution limitation when considering \textit{halo} clustering.  In particular, there are at least three relevant effects: halo exclusion, finite simulation volume, and the lower mass limit of the halo finder catalog. 
Halo exclusion implies there will be no clustering below the halo radius because the haloes will be touching, which will also impact the noise on small scales (see \citealt{Baldauf13} for more details).  In our analyses, however, we examine the clustering of subhaloes rather than FOF groups.  Subhaloes identified with \texttt{SUBFIND} can be in close proximity to each other and can even overlap spatially (particles are assigned to a single subhalo based on an energy unbinding criterion).  Thus, we do not anticipate halo exclusion effects to be relevant for our analyses.   With regards to simulation volume effects, the finite volume can result in biased halo number densities for very large haloes, since they are rare. For the majority of this study, however, we cut off haloes at mass $10^{12.5}h^{-1}$ M$_{\odot}$ which is a safe option even in moderate size boxes (see \citealt{vanDaalen+2014}).  Even without this cut, our results are likely to be insensitive to the rarest and most massive haloes (which, in BAHAMAS, are $\sim 10^{15}$ M$_\odot$), since there are orders of magnitude more haloes at lower masses which dominate the signal (see Fig.~\ref{fig:diff_mass_bins} below). 
Finally, with regards to the lower mass limit, we selected a lower cut off of $10^{11.5}h^{-1}$ M$_{\odot}$ corresponding to approximately 100 particles.  These are therefore relatively well-resolved haloes, in terms of being able to robustly estimate their masses and positions.  We have also verified that the BAHAMAS halo mass function follows that of other results in the literature (e.g., \citealt{Tinker2010}) down to this limit.


\section{Results}
\label{results}

In this section, we present our main results on the impact of baryon physics and neutrino free-streaming on the halo bispectrum.  We first examine the impact of baryons for the massless neutrino case in Section \ref{sec:baryons} and then we explore the effects of neutrinos and their degeneracy with baryonic effects in Section \ref{sec:neutrinos}.

\subsection{Impact of baryon physics}
\label{sec:baryons}

\begin{figure*} 
	\includegraphics[width=\columnwidth]{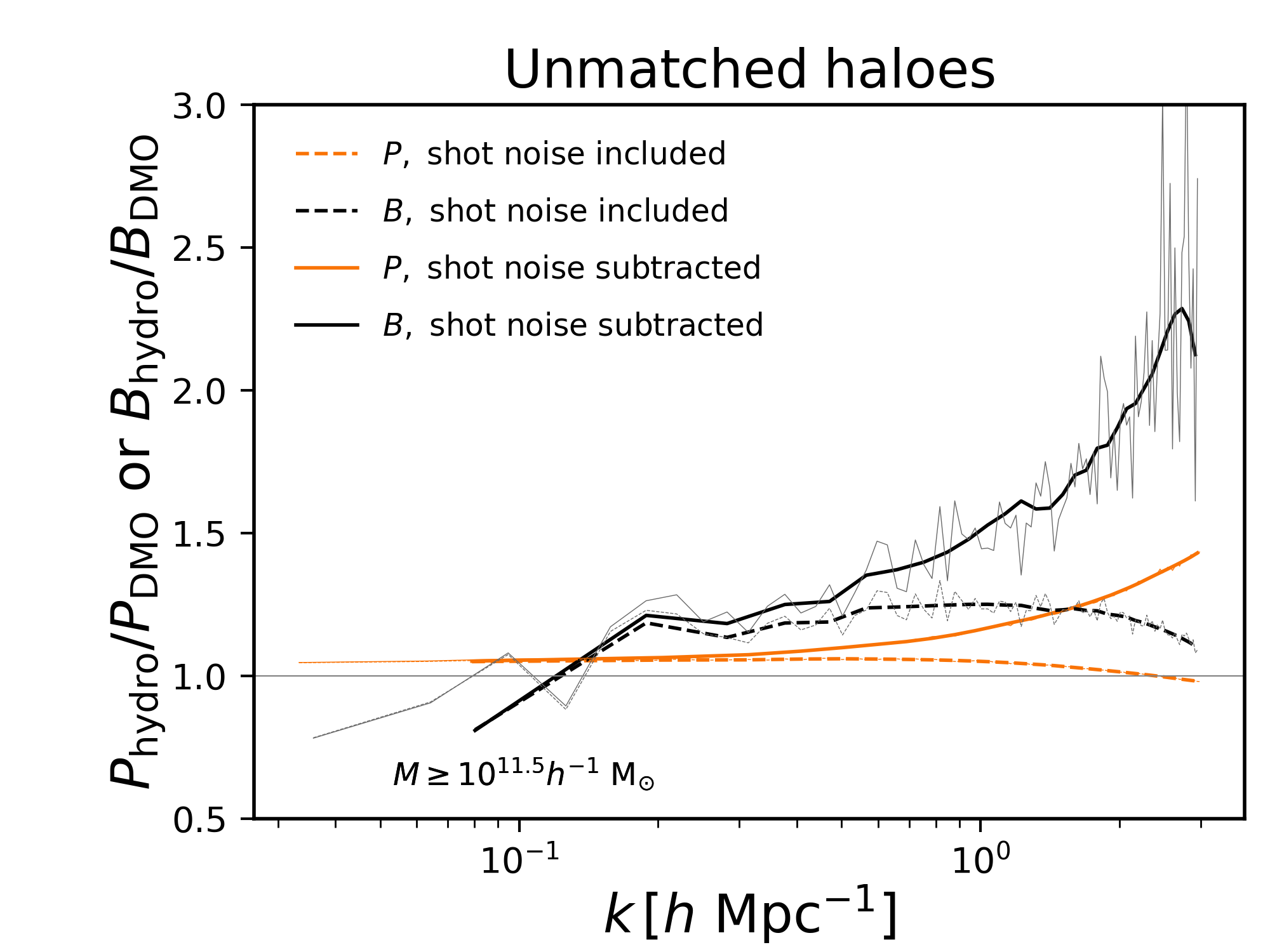}
    \includegraphics[width=\columnwidth]{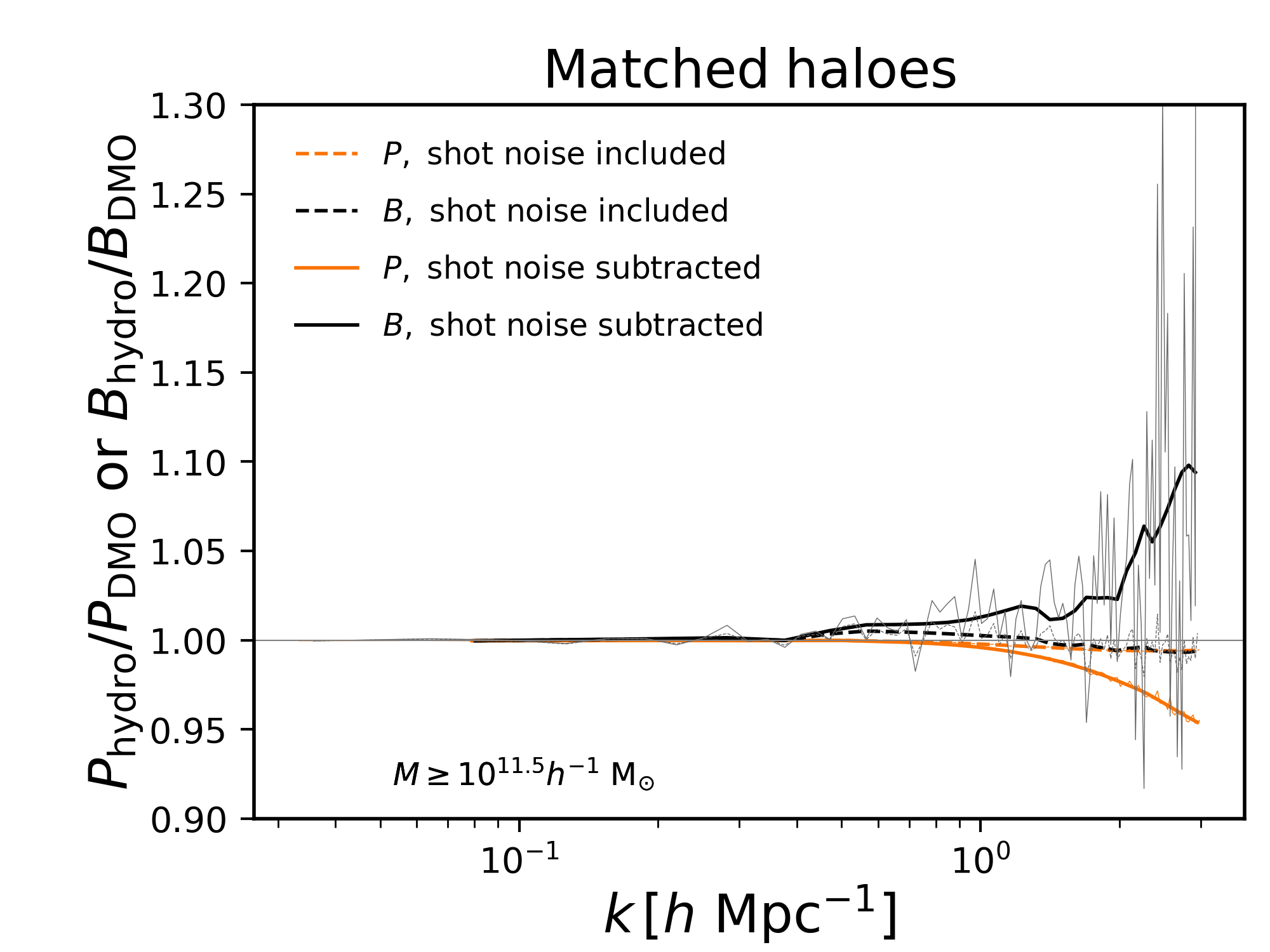}
  \caption{The ratio of the power spectrum (orange lines) and the bispectrum (black lines) measured from the fiducial hydro and DMO \texttt{BAHAMAS} simulations with (solid lines) and without (dashed lines) shot noise subtraction assuming equilateral triangles. All haloes (centrals and satellites) with masses exceeding $10^{11.5}h^{-1}$ M$_{\odot}$ are selected in the left panel (`unmatched' case). For the matched case (right panel), haloes are selected from the DMO case according to the mass criterion above and the same haloes are identified in the hydro run (regardless of their mass), hence a common set of haloes are used. The thick lines show the ratio with the smoothing procedure applied, while thin lines show the unsmoothed data. } 
  \label{fig:p_b_ratio_general} 
\end{figure*}

In Fig.~\ref{fig:p_b_ratio_general} we plot the ratio between the fiducial hydro and the DMO \texttt{BAHAMAS} simulations (both with massless neutrinos) of the halo power spectrum and bispectrum for equilateral triangles.  We consider both the raw spectra computed by \textsc{bskit} which includes a shot noise contribution and shot noise-subtracted spectra, where we have assumed the shot noise follows a Poisson distribution as given in equations (\ref{eq:snP}) and (\ref{eq:snB}). 
We note that when evaluating the shot noise contribution to the bispectrum in equation (\ref{eq:snB}), the power spectra should be shot noise-subtracted (according to eqn.~\ref{eq:snP}).  Since the power spectrum differs between the hydro and DMO cases, the bispectrum shot noise term will also, in general, be different for the hydro and DMO cases, even for a matched set of haloes.  Furthermore, it is clear that the bispectrum shot noise is not a constant (i.e., it depends on scale), unlike that for the power spectrum. As can be seen in Fig.~\ref{fig:p_b_ratio_general}, shot noise does not greatly affect large scales, but starting from $k \ga 0.1 \Mpch$ it begins to have an increasingly important effect.  Therefore, we will subtract the shot noise by default for the rest of the paper and we focus on the shot noise-subtracted results below.

Previous studies have shown that including galaxy formation physics in hydro simulations leads to a characteristic suppression in the non-linear matter power spectrum, which is due primarily to the ejection of baryons by AGN feedback  \citep[e.g.][]{vanDaalen11,Mummery17,Chisari19,Foreman+20, Stafford2020a, vanDaalen2020}. 
This effect was also shown to produce a suppression in the matter bispectrum by \citet{Foreman+20} (see also \citealt{Arico2021}). 
However, we show here that, at least at face value, haloes appear to show the opposite effect.
In particular, for the unmatched case, the `hydro' power spectrum and bispectrum are always greater than that of the corresponding `DMO' simulation, implying that haloes selected to be above a certain mass appear to be more clustered in a hydro simulation than in the corresponding collisionless one. Note that the apparent slight suppression in the bispectrum at large scales (small $k$) is due to sampling noise from finite box size effects. 

\begin{figure*} 
	\includegraphics[width=\columnwidth]{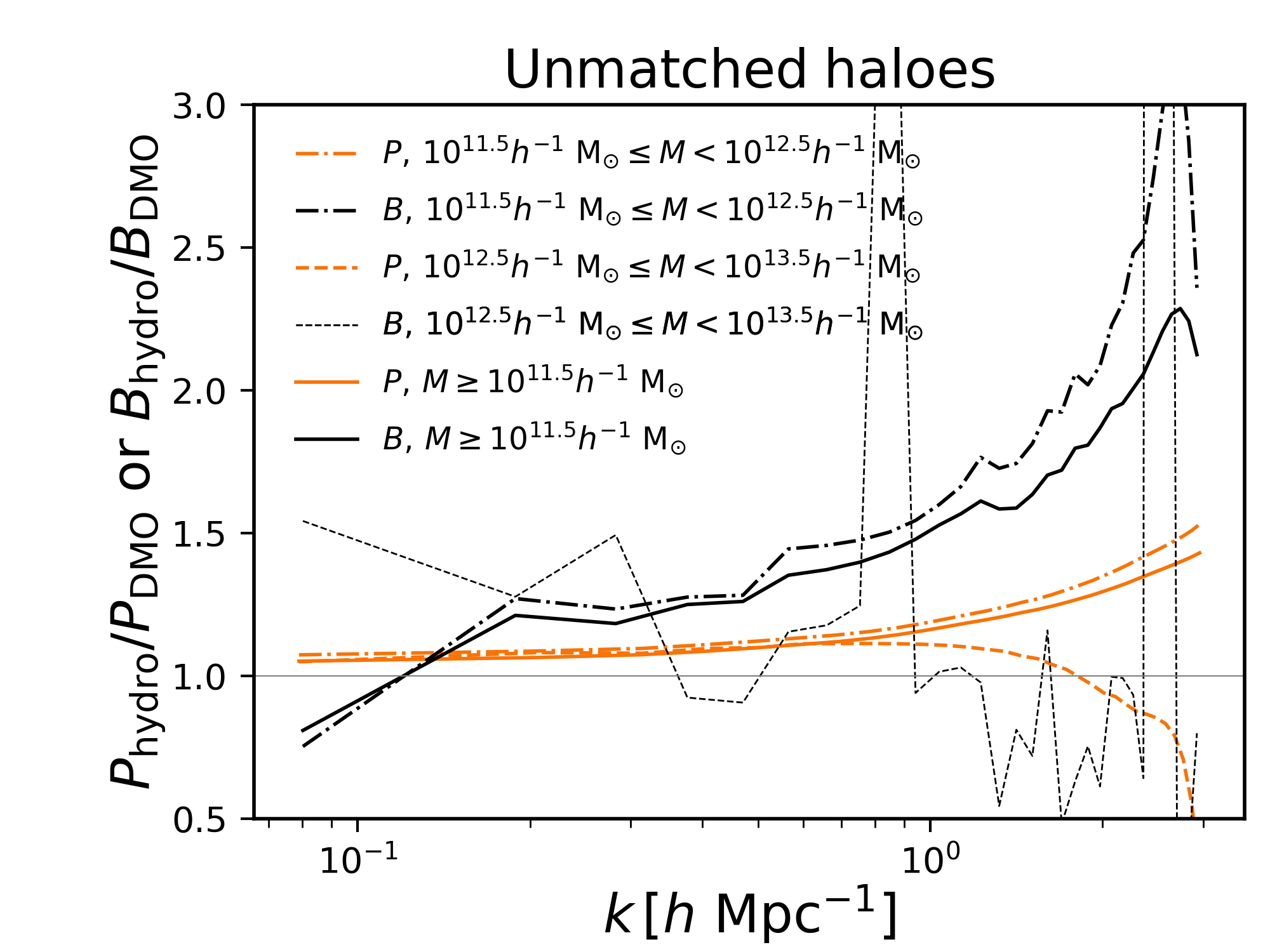}
\includegraphics[width=\columnwidth]
{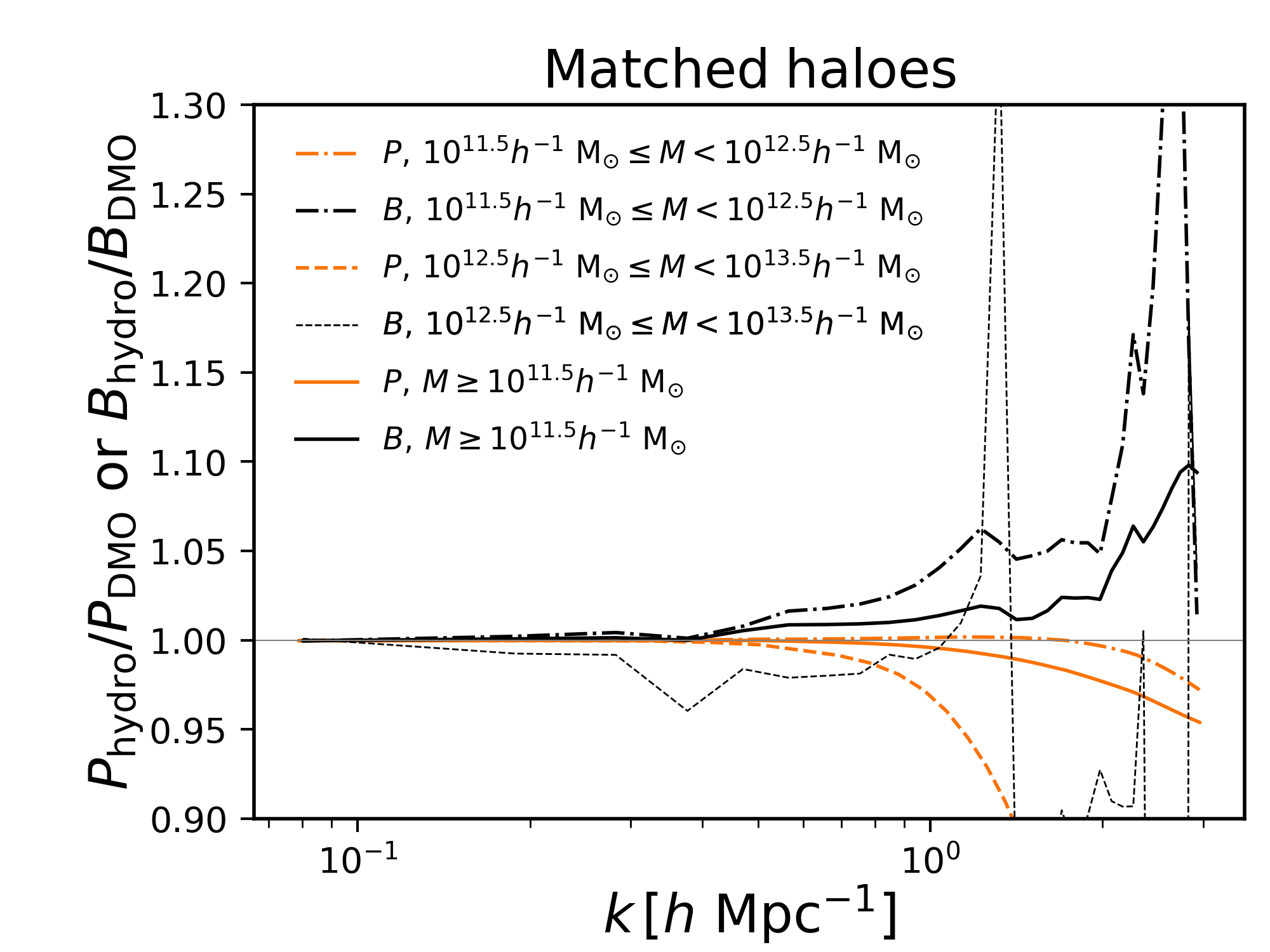}
  \caption{The ratio of the power spectrum (orange lines) and the bispectrum (black lines) measured from the hydro to DMO \texttt{BAHAMAS} simulations with shot noise subtraction for equilateral triangles. Different line styles denote different haloes mass cuts (solid lines demonstrate the same as at Fig.~\ref{fig:p_b_ratio_general}). The cut is done separately for hydro and DMO runs for the unmatched case. For the matched case, the cut is made according to DMO run for all lines. } 
  \label{fig:diff_mass_bins} 
\end{figure*} 

In the right panel of Fig.~\ref{fig:p_b_ratio_general} we consider the ratio of the hydro to the DMO case for a given (`matched') set of haloes common to the two simulations, rather than selecting haloes above a given mass limit (as in the left panel).  Here we see that on relatively large scales ($k \la 0.5$ $h$ Mpc$^{-1}$) the ratio now goes to unity, indicating no appreciable difference in the clustering of haloes in the two simulations on large scales. The same conclusion was made by \citet{vanDaalen+2014} when examining the 2-point correlation functions of a matched set of haloes in the \texttt{OWLS} simulations. 

As we are examining a common set of haloes in this comparison, it implies that the differences for the unmatched case (left panel) are primarily driven by a difference in the haloes that are selected when using a fixed mass cut in that case.  In particular, AGN feedback removes baryons from haloes in the hydro case, reducing their overall masses (e.g., \citealt{Cui2014,Cusworth2014,Velliscig2014,Schaller2015}).  Consequently, some haloes that would (in the absence of feedback) have been just above the mass cut are now below the threshold for selection and are therefore not selected.  The net result is that introducing baryon physics (feedback) results in the selection of intrinsically (in the absence of feedback) more massive systems.  Since more massive haloes are more strongly biased in their clustering with respect to the mean matter density (e.g., \citealt{Tinker2010}), the halo clustering is enhanced on large scales in the hydro simulation with respect to the DMO simulation when selecting haloes above a given mass limit.  For a common set of haloes (as in the right panel), however, there is no appreciable difference in the clustering on large scales.

\begin{figure*} 
	\includegraphics[width=\columnwidth]{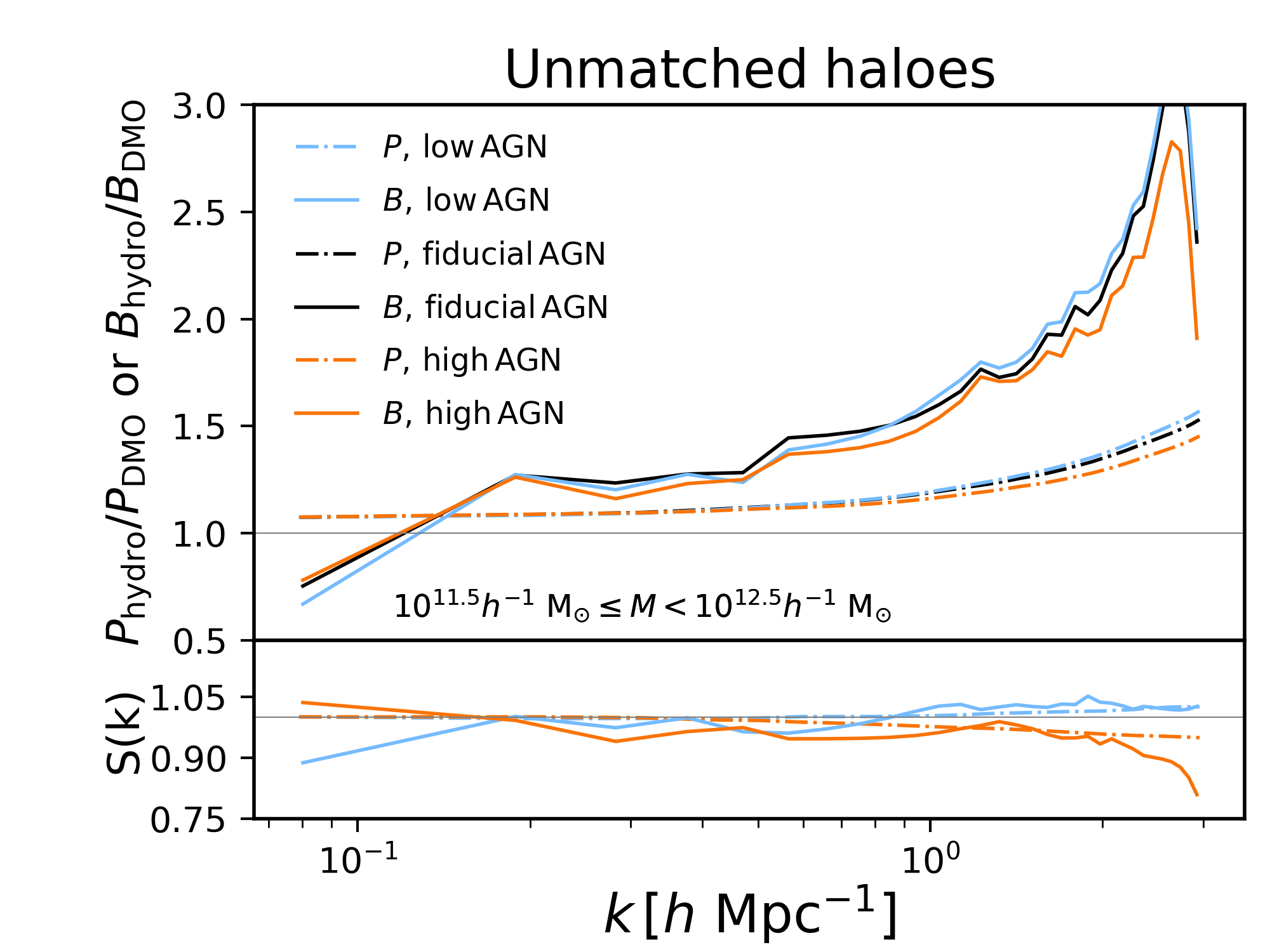}
\includegraphics[width=\columnwidth]
{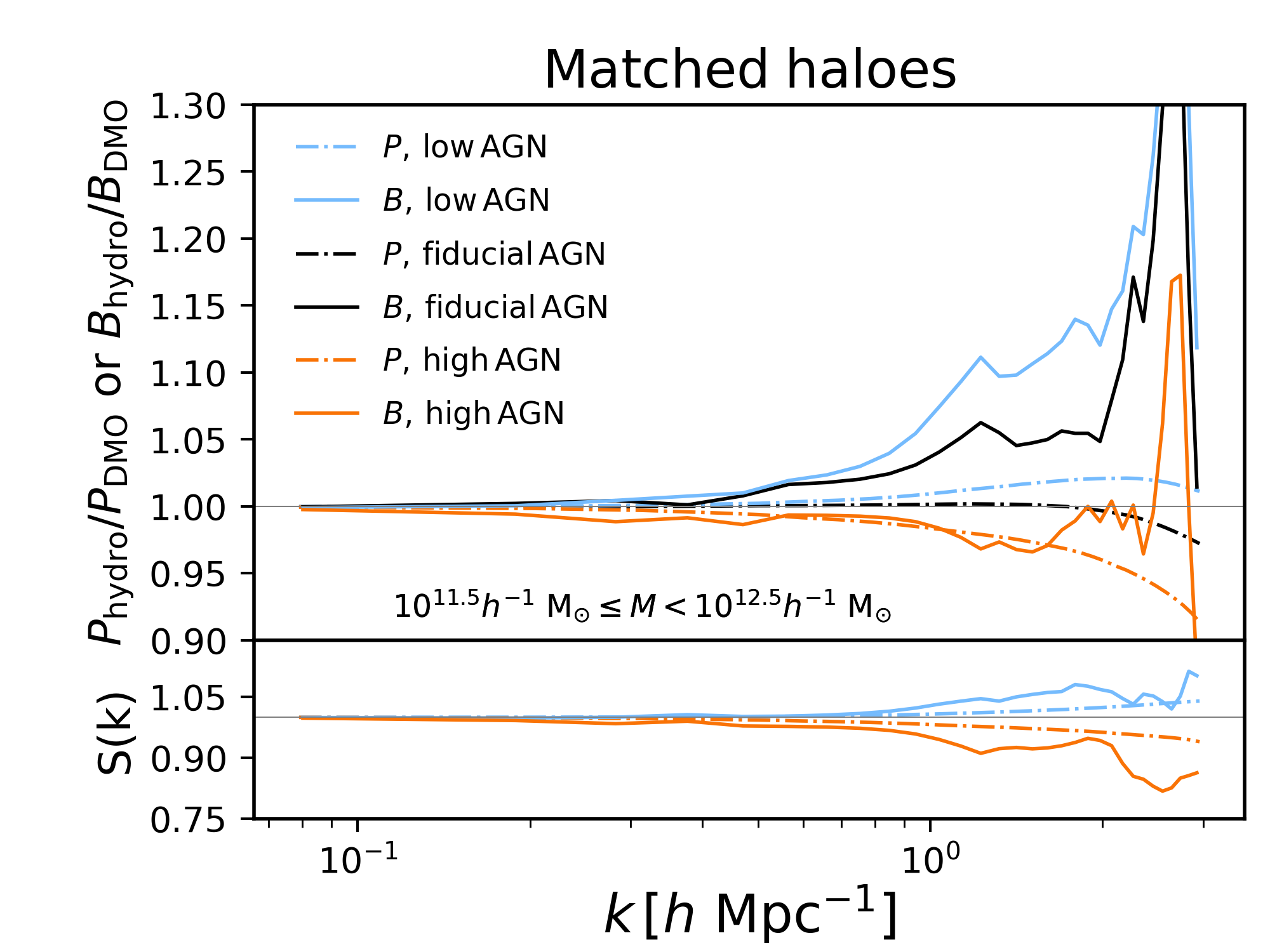}
  \caption{\textit{Upper panels}: The ratio of the power spectrum (dot-dashed lines) and the bispectrum (solid lines) measured from the hydro to DMO \texttt{BAHAMAS} simulations with shot noise subtraction for equilateral triangles for haloes masses in range $10^{11.5}h^{-1}$ M$_{\odot} \leq M < 10^{12.5}h^{-1}$ M$_{\odot}$. Different line colours represent variations the efficiency of AGN feedback. \textit{Lower panels}: The ratio $S(k)$ of the power spectrum and the bispectrum ratios between low/high AGN feedback models with respect to the fiducial (tuned AGN) model. }
  \label{fig:diff_agn} 
\end{figure*} 

On small scales ($k \ga 1$ $h$ Mpc$^{-1}$), however, the clustering is altered by hydrodynamics even for a common set of haloes.  In particular, we find that the power spectrum is slightly suppressed (at the 5\% level up to $k \approx 3$ $h$ Mpc$^{-1}$), whereas the bispectrum is enhanced (at the 10\% level up to $k \approx 3$ $h$ Mpc$^{-1}$).  On these scales, the clustering is almost certainly dominated by the so-called `1-halo' term, corresponding to satellites of massive haloes (groups and clusters).  Thus, deviations from unity are likely signifying differences in how the satellites are spatially distributed in massive haloes in the hydro case with respect to the DMO case.  Physically, such differences can potentially arise due to changes in the efficiency of tidal disruption, with gas ejection likely making satellites more susceptible to stripping at large (satellite-centric) scales, while star formation and adiabatic contraction would likely make disruption of the central regions of satellites more difficult to strip (with respect to the DMO case).  However, we cannot easily rule out that differences in the performance of the substructure finder (\texttt{SUBFIND}) between the DMO and hydro cases is also potentially playing a role here.  To test this, higher resolution simulations and/or alternative substructure finders (e.g., phase space-based finders) could be employed, but we leave this for future work.

\subsubsection{Halo mass dependence}

In the analysis presented above we explored the impact of baryon physics on the halo bispectrum at $z=0$ when selecting all haloes above $10^{11.5} h^{-1}$ M$_{\odot}$, corresponding roughly to a stellar mass selection of $M_* \ga 10^{10}$ M$_{\odot}$.  Here we attempt to examine the halo mass dependence of the results, noting that the effects of stellar and AGN feedback on the baryon fractions are expected to be halo mass dependent (e.g., \citealt{Schaller2015}).  We, therefore, expect the effects of feedback on the halo bispectrum to be sensitive to the halo mass range examined.  However, a limitation of the current study is that very large samples of haloes are required to accurately measure the bispectrum, thus breaking the sample up into multiple mass bins will yield less statistically robust results.  Nevertheless, we proceed here by examining the bispectrum (and power spectrum) in two adjacent mass bins of $10^{11.5}h^{-1}$ M$_{\odot} \leq M < 10^{12.5}h^{-1}$ M$_{\odot}$ and $10^{12.5} h^{-1}$ M$_{\odot} \leq M < 10^{13.5}h^{-1}$ M$_{\odot}$ and we compare these with our previous selection of $M \geq 10^{11.5}h^{-1}$ M$_{\odot}$ (Fig.~\ref{fig:diff_mass_bins}). Note that above the mass limit $10^{11.5}h^{-1}$ M$_{\odot}$ there are 721,459 haloes (or 100\%) in the DMO simulation and that the lower mass bin ($10^{11.5}h^{-1}$ M$_{\odot} \leq M < 10^{12.5}h^{-1}$ M$_{\odot}$) contains 638,832 of these haloes (or 88.6\%) while the higher mass bin contains only 76,016 haloes (or 10.5\%).

In Fig.~\ref{fig:diff_mass_bins} we show the ratio of the halo power spectra and bispectra (hydro to DMO) for the different halo mass selections for both the unmatched (left) and matched (right) cases.
The lower mass bin roughly follows the same trend as our total halo sample for the unmatched haloes, though the effects of baryons on both the power spectrum and bispectrum are slightly more pronounced in the low mass bin compared to the total halo selection.  The results for the higher mass bin ($10^{12.5} h^{-1}$ M$_{\odot} \leq M < 10^{13.5}h^{-1}$ M$_{\odot}$) are, unfortunately, too noisy to make any strong quantitative conclusions.  Larger simulation volumes are required to examine the halo bispectrum of high-mass objects.  Qualitatively, though, the results are consistent with a reduced impact due to baryon physics in comparison to the lower mass bin.  One can also infer this from the comparison of the total selection and lower mass bin comparison, since the latter is more strongly affected by baryons and the only difference in the selections is that the total case also contains high-mass haloes.

For the case of matched haloes (i.e., where haloes are selected according to some mass criterion in the DMO simulations and bijectively matched to the hydro simulation), shown in the right panel of Fig.~\ref{fig:diff_mass_bins}, we find much the same behaviour as in the right panel of Fig.~\ref{fig:p_b_ratio_general}.  Specifically, for a given set of haloes, baryon physics does not significantly alter the halo power spectrum or bispectrum on large scales ($k \la 0.5$ $h$ Mpc$^{-1}$) independent of halo mass.  On small scales (the 1-halo regime), however, the spectra are altered and the effects are largest for the lower mass bin (in agreement with the left panel) and we again find that the bispectrum is enhanced due to baryon effects whereas the power spectrum is suppressed (similar to the matter power spectrum).

\begin{figure*} 
	\includegraphics[width=\columnwidth]{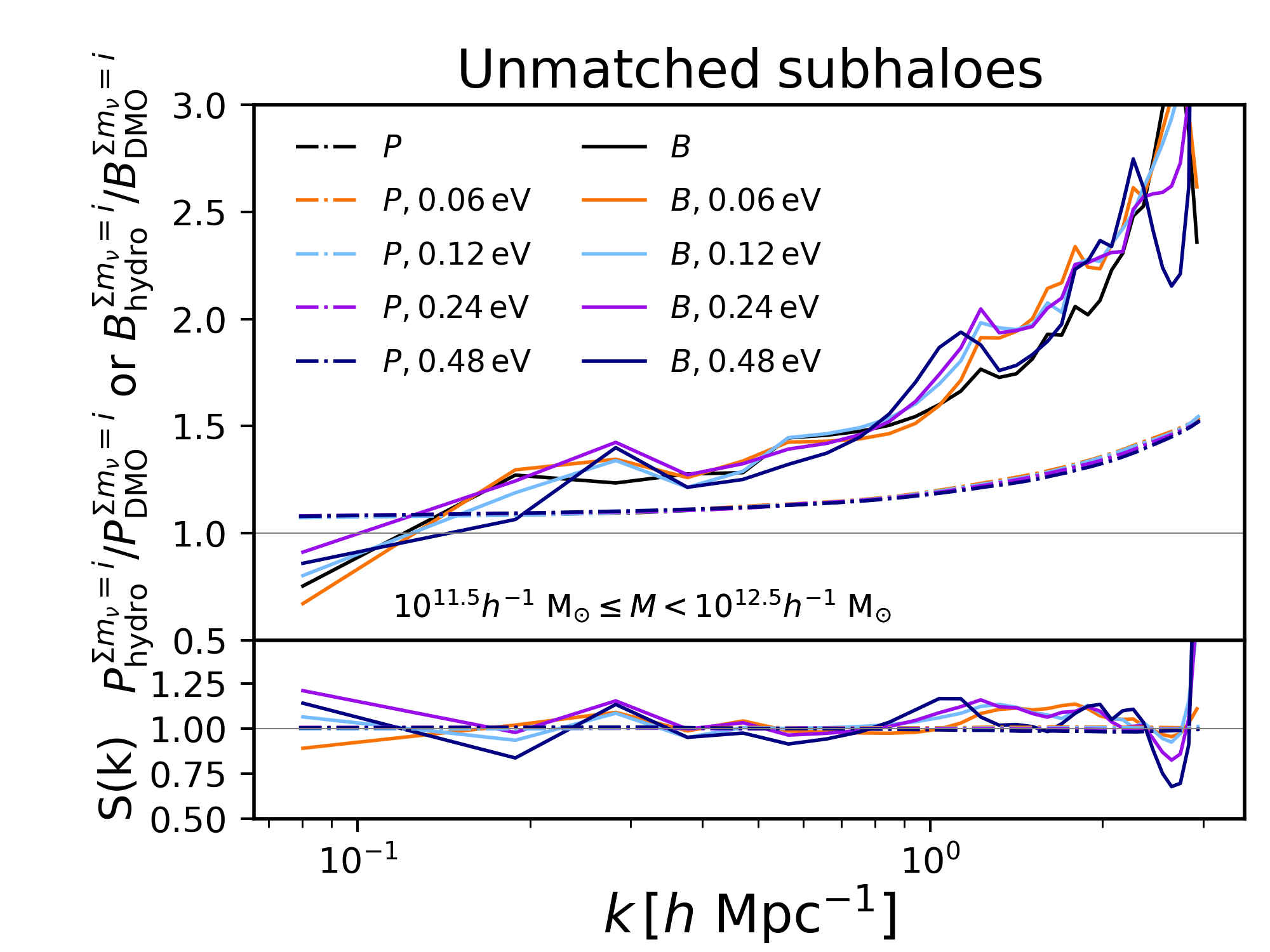}
\includegraphics[width=\columnwidth]
{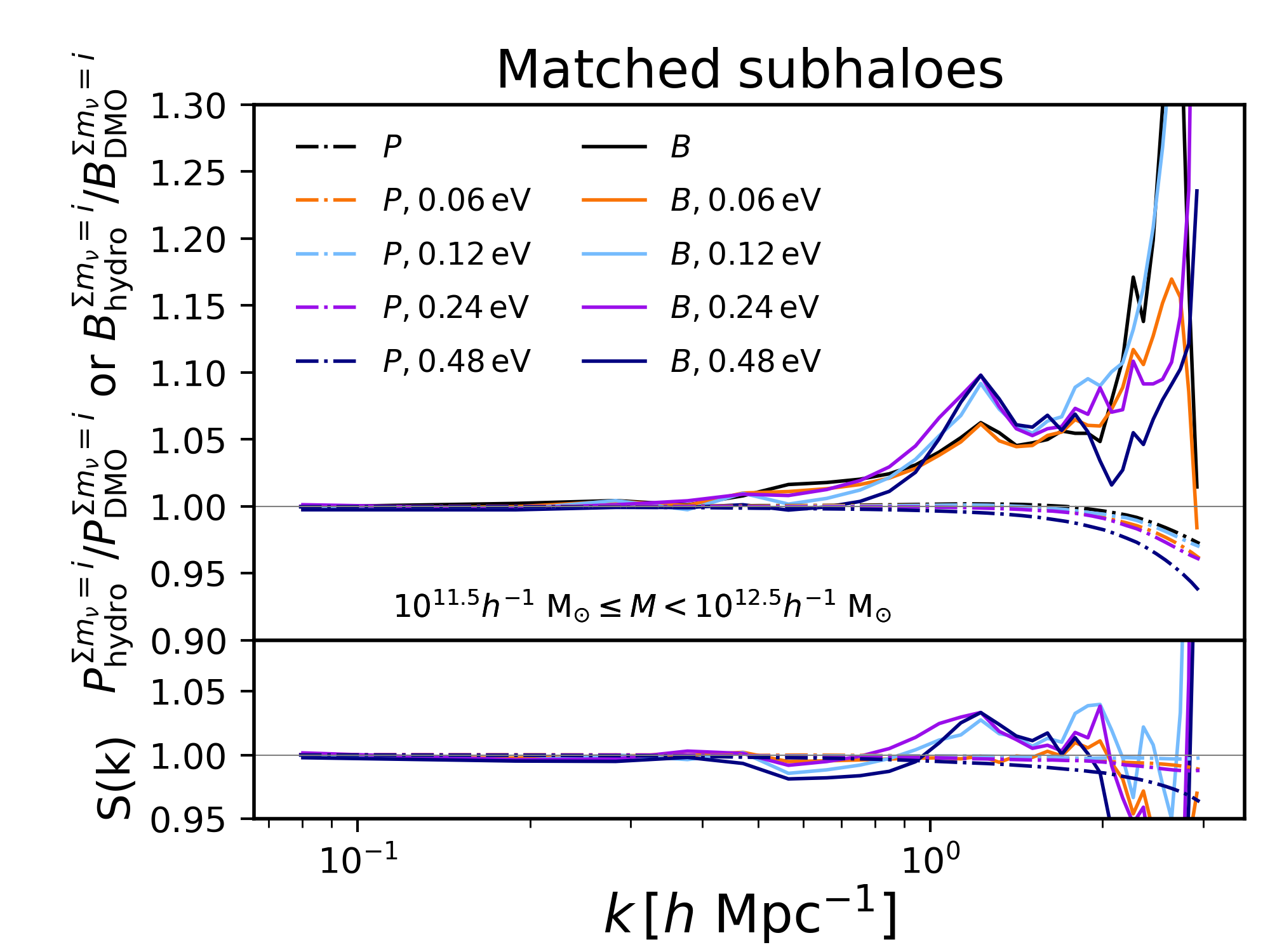}
  \caption{\textit{Upper panels}: The ratio of the power spectrum (dot-dashed lines) and the bispectrum (solid lines) measured from the hydro to DMO \texttt{BAHAMAS} simulations with shot noise subtraction for equilateral triangles for haloes masses in range $10^{11.5}h^{-1}$ M$_{\odot} \leq M < 10^{12.5}h^{-1}$ M$_{\odot}$. Different line colours represent different summed neutrino masses $\sum m_{\nu}$.  Note that the summed neutrino mass is varied simultaneously here for the hydro and DMO runs. \textit{Lower panels}: As in lower panels of Fig.~\ref{fig:diff_agn} but ratio $S(k)$ is of the massive neutrino cases with respect to the fiducial massless neutrino case.
  } 
  \label{fig:diff_nux} 
\end{figure*} 

\subsubsection{Feedback variations}
\label{sec:AGN}

In addition to the fiducial calibrated \texttt{BAHAMAS} model, \citet{McCarthy2018} introduced two additional simulations which varied the strength of the AGN feedback: `low AGN' and `high AGN'.  Specifically, they varied the subgrid AGN heating temperature parameter by $\pm0.2$ dex around the fiducial $\log_{10} [\Delta T_{\rm {heat}}] = 7.8$, which approximately brackets the upper and lower envelopes of the observed gas fractions of galaxy groups and clusters, as inferred from resolved X-ray observations (see fig.~3 of \citealt{McCarthy2018}).  Note that $\Delta T_{\rm {heat}}$ is the most important parameter in the model for determining the gas fractions of groups/clusters.  Higher values of $\Delta T_{\rm {heat}}$ correspond to more energetic (and more bursty) AGN feedback episodes and vice-versa.

We show in Fig.~\ref{fig:diff_agn} the effect of varying $\Delta T_{\rm {heat}}$ on the halo power spectrum and bispectrum for a halo mass selection of $10^{11.5}h^{-1}$ M$_{\odot} \leq M < 10^{12.5}h^{-1}$ M$_{\odot}$, once again through the ratio of these statistics measured from the hydro simulations computed with respect to the DMO case.
For the unmatched case, the power spectrum and bispectrum ratios for all three AGN models show an enhancement due to feedback.  Interestingly, it is the `low AGN' model which shows the largest effect particularly on small scales, whereas for the \textit{matter} power spectrum and bispectrum the opposite is true (e.g., fig.~13 of \citealt{Foreman+20}, see also \citealt{vanDaalen2020}).  In the case of the matter clustering, the interpretation is straightforward: more energetic feedback leads to increased baryon ejection which reduces the overall matter clustering.  For the halo clustering, particularly on small scales which are dominated by the 1-halo term (satellites of groups/clusters), it is plausible that particularly energetic feedback leads to satellites being more susceptible to environmental processing (tidal heating/stripping), reducing the power on small scales compared to a run with less energetic feedback.  It is interesting that in the matched halo case (right panel of Fig.~\ref{fig:diff_agn}) the fiducial and low AGN runs show enhanced halo bispectra (and power spectra, for low AGN) relative to the DMO case, whereas the high AGN shows suppressed halo clustering.  This suggests there is a relatively fine line between baryon effects enhancing the clustering on small scales relative to DMO (e.g., due to star formation and adiabatic contraction of the satellites subhaloes) and enhanced tidal stripping due to baryon ejection.  

The upshot of the sensitivity of the halo clustering due to baryon effects is that the simulations probably cannot be used to robustly predict whether there should be an enhancement or a suppression relative to DMO simulations on small scales (for a common set of haloes), as the answer appears to depend sensitively on the balance between feedback and environmental physics.  
This is in contrast to the findings of \citet{Foreman+20} (see, e.g., their fig.~13), in which the matter bispectra of the three BAHAMAS hydro simulations consistently show a suppression relative to the DMO case, whereas our measurements of the \textit{halo} bispectrum show either a suppression or an enhancement depending on the AGN feedback strength. This potentially complicates the interpretation of observational measurements, since a probe such as the cosmic shear bispectrum would observe a suppression while the galaxy 3-point function might be suppressed or enhanced. A joint analysis of the matter and halo bispectra would therefore be useful for simultaneously constraining the effects of feedback and environmental physics on small scales.  In this regard, it is noteworthy that the effects are typically much stronger in the bispectrum than in the power spectrum (consistent with the findings of \citealt{Foreman+20}), suggesting that future measurements of the small-scale bispectrum are a promising tool for studying these effects.

\subsection{Impact of neutrino free-streaming}
\label{sec:neutrinos}

Of the various possible extensions to the standard $\Lambda$CDM model, massive neutrinos are perhaps the most well motivated, as the results of atmospheric and solar oscillation experiments imply that the three active species of neutrinos have a \textit{minimum} summed mass, $\Sigma m_{\nu}$, of at least 0.06 eV (0.1 eV) when adopting a normal (inverted) hierarchy (see \citealt{Lesgourgues2006} for a review).  At present, cosmological studies place the strongest upper limits on the summed neutrino mass.  Galaxy clustering studies, in particular, have a rich history of placing important constraints on the summed neutrino mass (e.g., \citealt{Tegmark2004,Cole2005,Sanchez2006,Beutler2014_nu,Abbott2018}), with its sensitivity coming from the fact that neutrinos suppress the growth of structure (i.e., reduce the clustering amplitude relative to a massless neutrino cosmology) on scales smaller than the associated free-streaming scale.  In the present study, we use the massive neutrinos extension of \texttt{BAHAMAS} \citep{Mummery17,McCarthy2018} to explore the joint effects of baryon physics and neutrinos, and their possible degeneracy, on the halo power spectrum and bispectrum.  In addition to the massless neutrino cases explored above, \texttt{BAHAMAS} has four variations of summed neutrino mass, with $\Sigma m_{\nu} = 0.06$,  $0.12$, $0.24$ or $0.48$ eV (Table~\ref{tab:sims}).  As mentioned in Section \ref{BAHAMAS}, neutrinos are implemented in the simulations using the semi-linear method of \citet{Bird2013}, which is sometimes referred to as a `linear response' method.  For further details of the implementation within \texttt{BAHAMAS}, we refer the reader to \citet{McCarthy2018}.

\begin{figure*} 
	\includegraphics[width=\columnwidth]{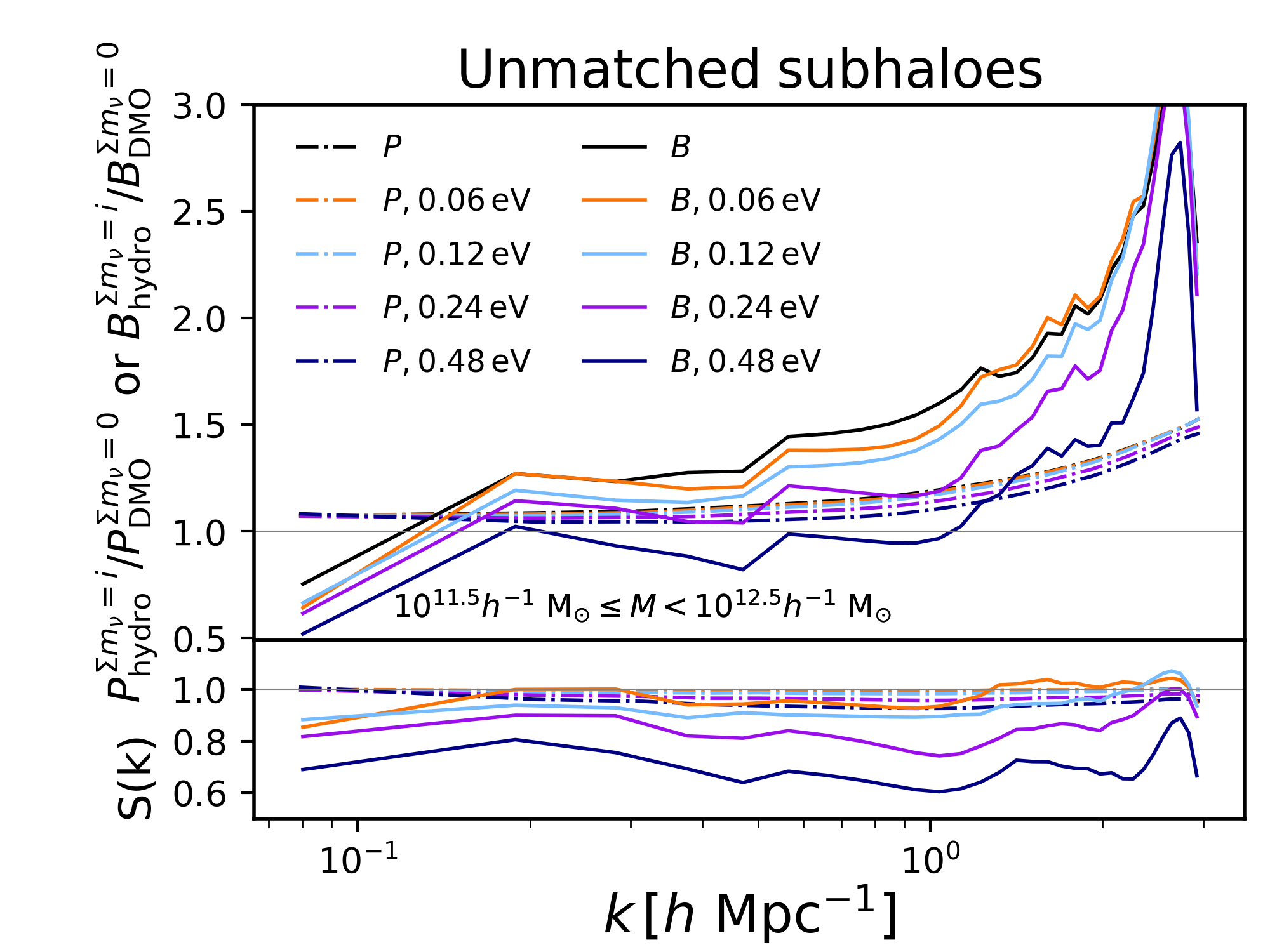}
\includegraphics[width=\columnwidth]
{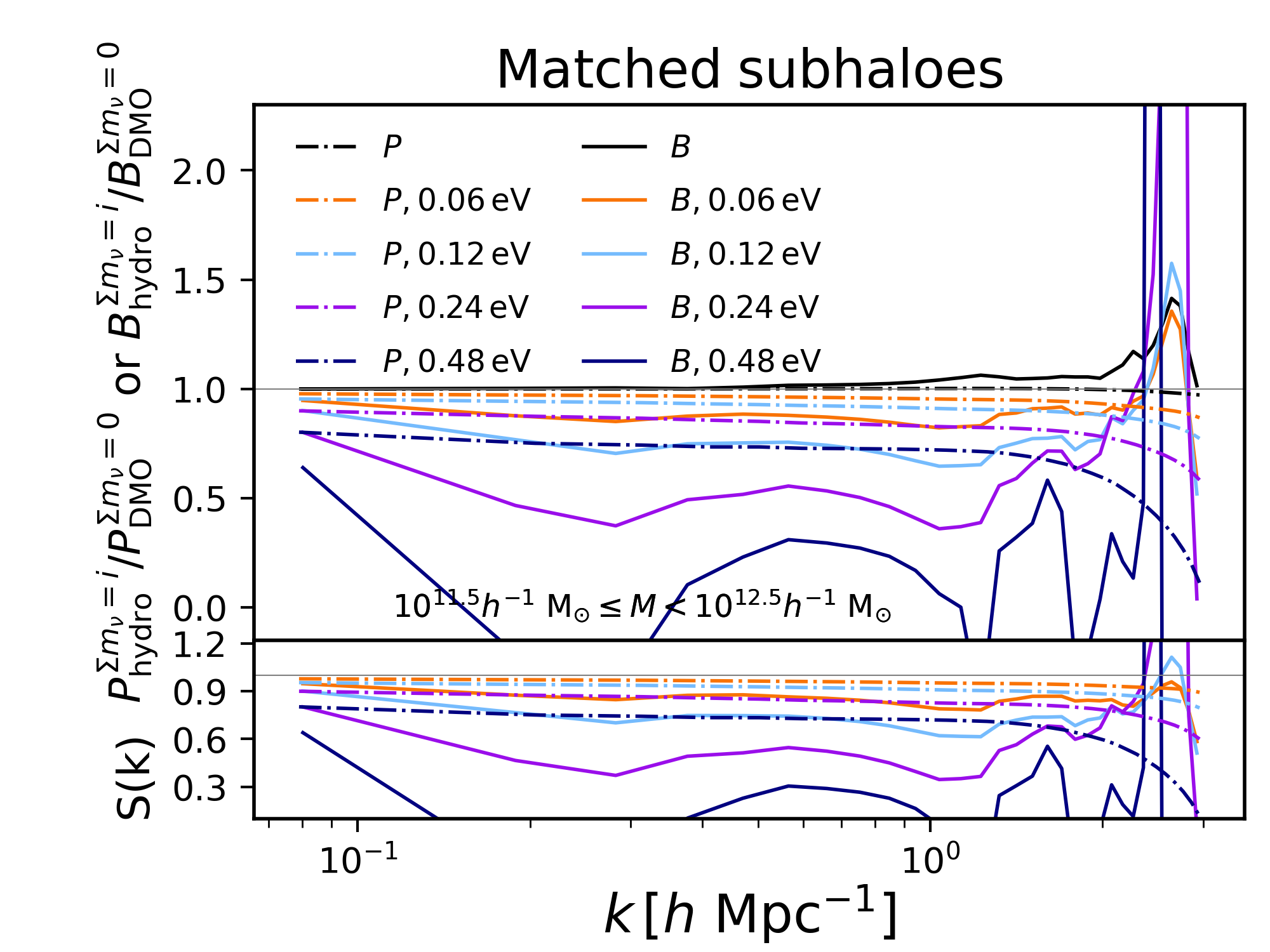}
  \caption{As in Fig.~\ref{fig:diff_nux} except here the DMO case corresponds to massless neutrinos.  Thus, both baryon physics and neutrinos will alter the ratio of the hydro simulation with respect to the DMO run.
} 
  \label{fig:diff_nu0_nux} 
\end{figure*} 

In Fig.~\ref{fig:diff_nux} we show the ratio of the power spectrum and the bispectrum (hydro to DMO) where the summed neutrino mass is systematically varied for both the hydro and DMO runs (simultaneously).  In essence, this comparison explores whether the relative impact of baryon physics on the halo clustering is dependent on the summed neutrino mass.  If the ratio varies significantly between the simulations, this implies that the impact of baryon physics is dependent on the summed neutrino mass, whereas if the ratio does not vary between the simulations then the effects of baryon physics are independent of (i.e., separable from) those of massive neutrinos.  As shown in Fig.~\ref{fig:diff_nux} (focusing on the bottom subpanels, in particular, which show the ratio with respect to the massless neutrino case) there is no evidence for a strong systematic dependence on the summed neutrino mass, with the results being very similar to the massless case.  The noise in the halo bispectrum measurements means we can only place an upper limit on the potential effect but, for the matched case, the impact of baryons on the halo bispectrum agrees amongst all of the neutrino simulations to typically a few per cent accuracy (and to higher accuracy for the halo power spectrum).  Thus, the impact of baryons and neutrinos appear to be separable effects for the halo power spectrum and bispectrum, as previously found for the matter clustering in \citet{Mummery17} and \citet{vanDaalen2020}.

In Fig.~\ref{fig:diff_nu0_nux} we explore the combined effects of baryon physics and neutrino free-streaming on the halo clustering, by taking the ratio of the hydro simulations of varying neutrino mass to the DMO run with \textit{massless} neutrinos.  Thus, both feedback and neutrinos will impact this metric.  In fact, there are four effects present when considering the left panel of Fig.~\ref{fig:diff_nu0_nux} (unmatched haloes); as both feedback and free-streaming affect the halo mass (i.e., alter the halo selection) and both feedback and free-streaming affect the clustering (although we have shown that the clustering is only affected on small scales by baryon physics after we control for the halo mass alteration).  Neutrinos alter the halo mass because they represent a form of dark matter that is effectively unable to collapse, whereas all of the dark matter can collapse in the massless neutrino case.  \citet{Mummery17} have shown that it is the largest haloes (massive clusters) that are most strongly affected in terms of halo mass.

In the left panel of Fig.~\ref{fig:diff_nu0_nux} we see that the halo power spectra and bispectra are typically enhanced with respect to the DMO with massless neutrinos case.  Increasing the summed neutrino mass tends to reduce this enhancement, implying that the free-streaming reduction in clustering amplitude is a more significant effect than the reduction in halo mass that is due to neutrinos.  However, for this range of neutrino masses, the change in halo mass due to baryon ejection is sufficient to overcome the impact of neutrinos, which is why there is an overall enhancement in the clustering rather than a suppression.  The one exception to this is the $\Sigma m_{\nu} = 0.48$ eV case, which shows a slight suppression on quasi-linear scales.  Here the neutrino suppression is sufficiently strong to overcome the enhancement due to baryon ejection.

In the right panel of Fig.~\ref{fig:diff_nu0_nux} we do the same comparison but now for a common set of haloes, selected from the DMO massless neutrino run and bijectively matched to each of the hydro simulations with varying neutrino masses.  As explained before, by examining a common set of haloes we are removing the impact of changes in halo mass, due to both feedback and neutrino free-streaming, on the halo selection.  Here we see a strong suppression in the clustering which is present on virtually all scales plotted, apart from the very smallest scales where the enhancement due to baryon physics begins to dominate.  In fact, the suppression will be present out to approximately the free-streaming scale, which corresponds to $k_{\rm fs} \approx 0.05 \ \Mpch$ for $\Sigma m_{\nu} =0.06$ eV (see eqn.~9 of \citealt{Bird2013}).  In common with the impact of feedback, we find that the bispectrum is affected much more significantly than is the power spectrum, although we note that estimates of the bispectrum are more affected by noise (sample variance) than is the power spectrum.  In principle, though, upcoming precision measurements of the bispectrum may be able to place strong constraints on both cosmology and models of galaxy formation simultaneously, by using small scales to probe feedback and environmental processing and large scales to help constrain the neutrino mass.  Certainly the combination of the power spectrum and bispectrum will allow for much more powerful constraints than may be obtained from the power spectrum alone.

Finally, in the above discussion we have neglected the potential degeneracies between the summed neutrino mass, $\sigma_8$, and baryons.  It is known, for example, that using clustering data alone which is restricted to quasi-linear and non-linear scales (e.g., as in current cosmic shear surveys) that $\Sigma m_{\nu}$ and $\sigma_8$ can be strongly degenerate.  Because the BAHAMAS neutrino simulations are `CMB normalised', varying $\Sigma m_{\nu}$ leads to a variation in $\sigma_8$ as well.  To explore the degeneracy in detail would require additional simulations which hold $\sigma_8$ fixed as the neutrino mass is varied (or vice-versa), which is beyond the scope of the present study.  We note, however, that with the increasing quality of weak lensing and spectroscopic galaxy clustering surveys, one can exploit the fact that neutrino free-streaming, variations in $\sigma_8$, and the impact of baryons on the matter and halo clustering all have different scale and redshift dependencies (see, e.g., \citealt{Mummery17} who compared the scale and redshift dependencies of baryons and massive neutrinos on various clustering statistics).  We therefore expect upcoming clustering studies to be able to place much stronger constraints on these effects than is currently possible.

\section{Summary and conclusions}
\label{conclusions}

The primary goal of this paper was to demonstrate the power of the halo bispectrum as a tool for testing cosmological models and galaxy formation physics (particularly AGN feedback). 
To this end, we measured the halo power spectrum and the halo bispectrum from the \texttt{BAHAMAS} hydro and dark matter simulation suites.  Aside from the fiducial calibrated \texttt{BAHAMAS} hydro simulation (which was calibrated to reproduce the galaxy stellar mass function and the gas fractions of galaxy groups/clusters), the suite contains runs which vary the strength of the AGN feedback and the summed mass of neutrinos, allowing us to explore the sensitivity of the halo clustering to these effects.  We modified the publicly available matter clustering code \textsc{bskit} \citep{Foreman+20} to measure the halo clustering in \texttt{BAHAMAS}.

The main results of our study may be summarised as follows:

\begin{itemize}

\item The bispectrum measured from the simulations in general depends on the adopted triangular configuration.  It is common place to adopt an equilateral configuration mostly for convenience.  We explored the dependence of our main results (the relative impact of baryon physics on the halo bispectrum) on the adopted triangular configuration in Fig.~\ref{fig:diff_tri_config} (see also Appendix \ref{App:tri_choice}), finding that the equilateral choice tends to maximise the impact of baryons.  We adopted the equilateral choice for the remainder of the analysis but note that the results are somewhat sensitive to the choice of triangular configuration.  In the future, it may be possible to exploit this sensitivity as a further test of baryon and cosmological physics.

\item When selecting haloes above a given mass, baryon physics (particularly AGN feedback) tends to \textit{enhance} the halo power spectrum and bispectrum relative to a collisionless (dark matter-only, DMO) case (see left panel of Fig.~\ref{fig:p_b_ratio_general}).  This enhancement is driven by the change in halo mass due to AGN feedback.  Specifically, haloes just above the selection threshold in the DMO case have their masses reduced due to gas ejection and therefore not selected in the hydro simulation case.  Consequently, the mean bias of haloes selected in the hydro case is larger than that of the DMO case, giving rise to an apparent enhancement in the clustering.  Comparing the clustering of a \textit{common set} of haloes in the hydro and DMO cases (see right panel of Fig.~\ref{fig:p_b_ratio_general}) demonstrates that baryon physics does not appreciably affect the clustering on large scales of $k \la 0.5$ $h$ Mpc$^{-1}$.  On small scales, dominated by the 1-halo term (satellites of massive groups and clusters), the power spectrum and bispectrum can be significantly affected ($\approx 10\%$ level) and, interestingly, we find for the fiducial calibrated \texttt{BAHAMAS} model that the bispectrum is enhanced with respect to the DMO case, while the power spectrum is suppressed.

\item We find that the effects described above are a function of halo mass, with lower masses ($\sim10^{12}$ M$_\odot$) being more significantly affected by baryon physics (see Fig.~\ref{fig:diff_mass_bins}).

\item We explored the sensitivity of the results to variations in galaxy formation physics by using two variations from the \texttt{BAHAMAS} suite (`low AGN', `high AGN').  Interestingly, we find that reducing the efficiency of AGN feedback results in an increased enhancement in the halo bispectrum on small scales (see right panel of Fig.~\ref{fig:diff_agn}), which is plausibly due to the fact satellites are more resilient against tidal disruption in this case.  Increasing the efficiency of AGN feedback can actually change the sign of the effect, such that the halo bispectrum is suppressed with respect to the DMO case.  This suggests that there is a very fine balance between star formation/adiabatic contraction (which tend to make haloes less susceptible to tidal heating/stripping) and gas ejection (which make them more susceptible).  Small-scale measurements of the halo bispectrum, therefore, offer a means of simultaneously probing feedback and environmental physics.

\item Using the massive neutrinos extension of \texttt{BAHAMAS} we explored the joint effects of neutrino free-streaming and baryon physics (and their potential degeneracy) on the halo clustering.  We showed that the effects of baryon physics are largely independent of (separable from) the effects of neutrinos (see Fig.~\ref{fig:diff_nux}).  For a common set of haloes, we find neutrinos strongly suppress the halo bispectrum on all but the smallest scales probed in the present study (the smallest scales are dominated by baryon physics which can either enhance or suppress the bispectrum, as discussed above).  See the right panel of Fig.~\ref{fig:diff_nu0_nux}.  
\end{itemize}

Our results have demonstrated that the halo power spectrum and particularly the halo bispectrum are highly sensitive to neutrino effects on the largest scales and baryon physics (AGN feedback and star formation) on the smallest scales.   This bodes well for upcoming precision measurements of the bispectrum for surveys like Euclid, LSST, and DESI.  To capitalise on these measurements and the implied sensitivity what is required now is a large suite of very large volume cosmological simulations that systematically and simultaneously explore variations in cosmological parameters and galaxy formation physics.  This is the aim of ongoing work.

\section*{Acknowledgements}
The authors thank the referees for constructive suggestions which have improved the paper.
They also thank Andreea Font, Jaime Salcido, and Amol Upadhye for helpful feedback on the paper and acknowledge helpful discussions with Davit Alkhanishvili, William Coulton, Alexander Eggemeier, ChangHoon Hahn, Andrea Oddo, and Joop Schaye. 
This project has received funding from the European Research Council (ERC) under the European Union's Horizon 2020 research and innovation programme (grant agreement No 769130). SGS acknowledges an STFC doctoral studentship.
This work used the DiRAC@Durham facility managed by the Institute for Computational Cosmology on behalf of the STFC DiRAC HPC Facility. The equipment was funded by BEIS capital funding via STFC capital grants ST/P002293/1, ST/R002371/1 and ST/S002502/1, Durham University and STFC operations grant ST/R000832/1. DiRAC is part of the National e-Infrastructure.

\section*{Data availability}
The data underlying this article may be shared on reasonable request to the corresponding author.




\bibliographystyle{mnras}
\bibliography{yankelevich} 

\appendix

\section{Halo mass-weighted spectra}
\label{App:weight}

\begin{figure*} 
	\includegraphics[width=\columnwidth]{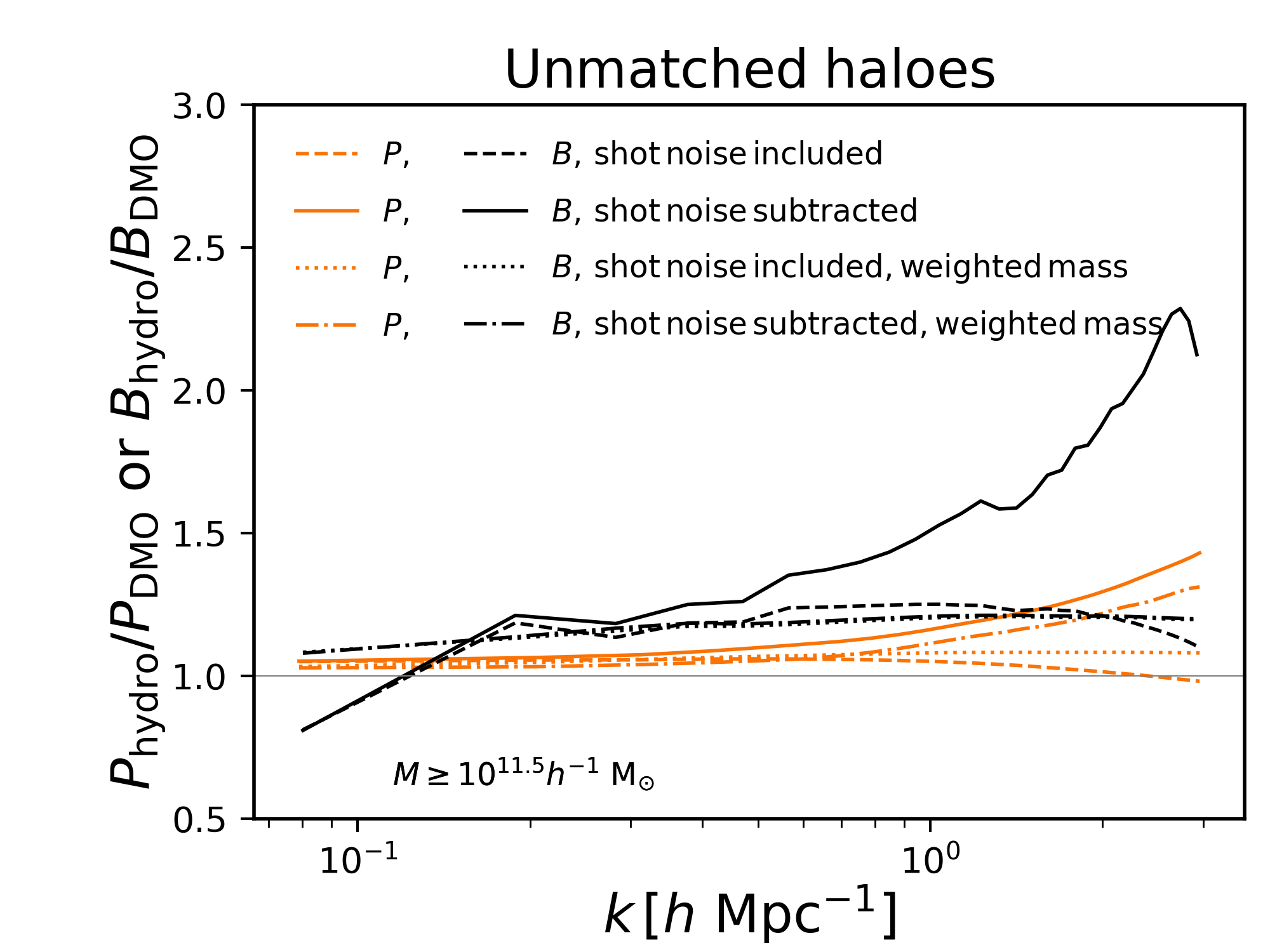}
    \includegraphics[width=\columnwidth]{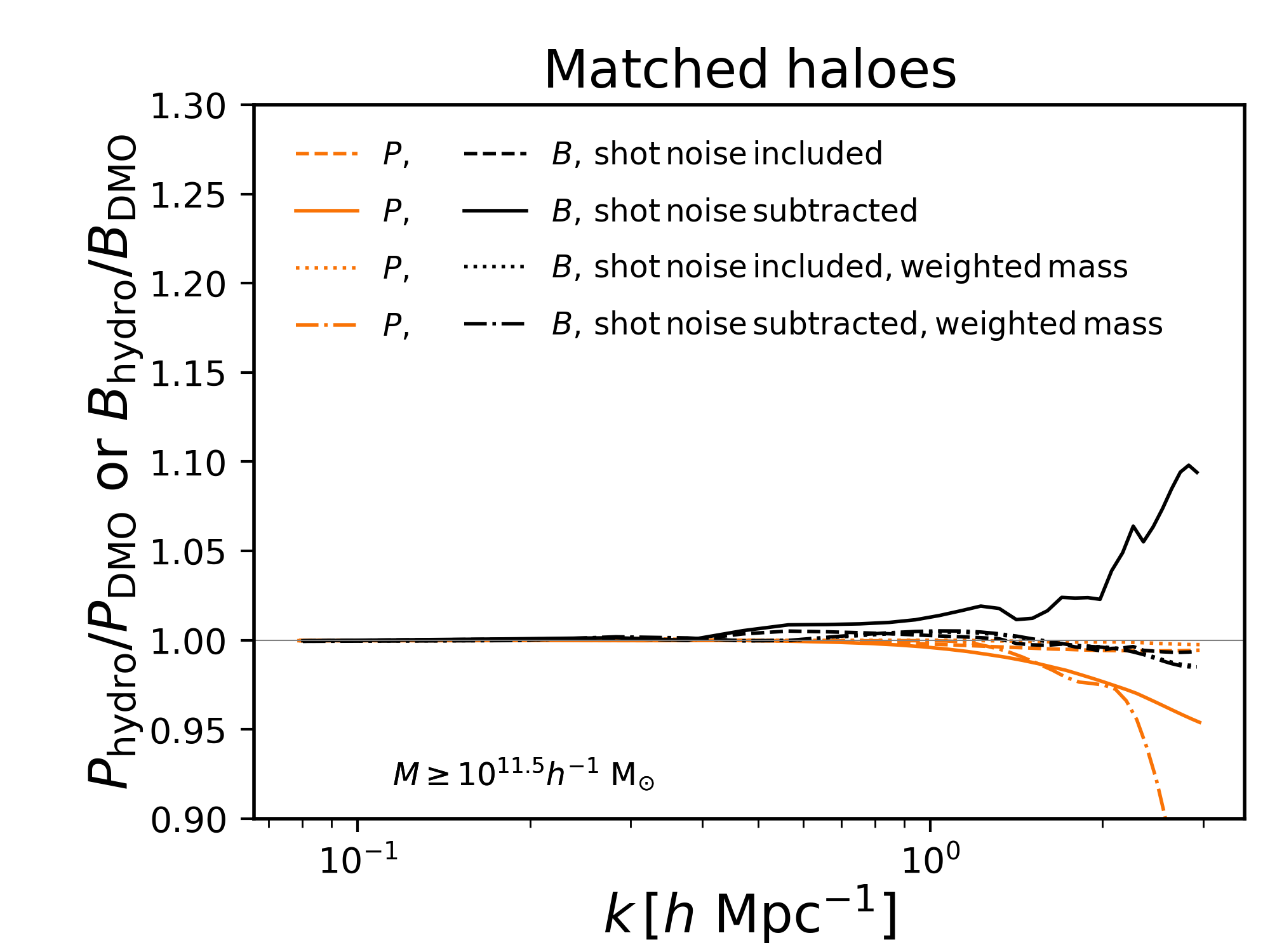}
  \caption{The same as at Fig.~\ref{fig:p_b_ratio_general} but including cases where the haloes are weighted by their masses when computing the power spectra and bispectra.} 
  \label{fig:weight_mass} 
\end{figure*} 

In the main analysis, when computing the halo power spectra and bispectra we treat all haloes equally (same weighting).  Here we explore an alternative weighting scheme, which is to weight each halo by its mass.  It was shown previously by \citet{Seljak-Hamaus-Desjacques09} that the effects of shot noise can be significantly reduced in the halo power spectrum by adopting such a weighting scheme.   In \textsc{bskit} there is an option to include particle weight into analysis when computing the grids.  We simply substitute in the halo mass. For the matched case we used DMO halo masses for both DMO and hydro grids. 

We demonstrate our findings in Fig.~\ref{fig:weight_mass}.  While we see no obvious advantage in terms of the shot noise for the bispectrum, it is interesting to note that the effects of baryon physics are reduced significantly when weighting haloes by their mass instead of equally weighting each halo.  This is likely because higher mass haloes are less significantly affected by ejective feedback.  Thus, if the aim is to mitigate against the effects of baryons, employing a halo mass weighting scheme goes some way towards achieving this objective.

\section{Triangular configuration dependence as a function of scale}
\label{App:tri_choice}

\begin{figure*}
	\includegraphics[width=14cm]{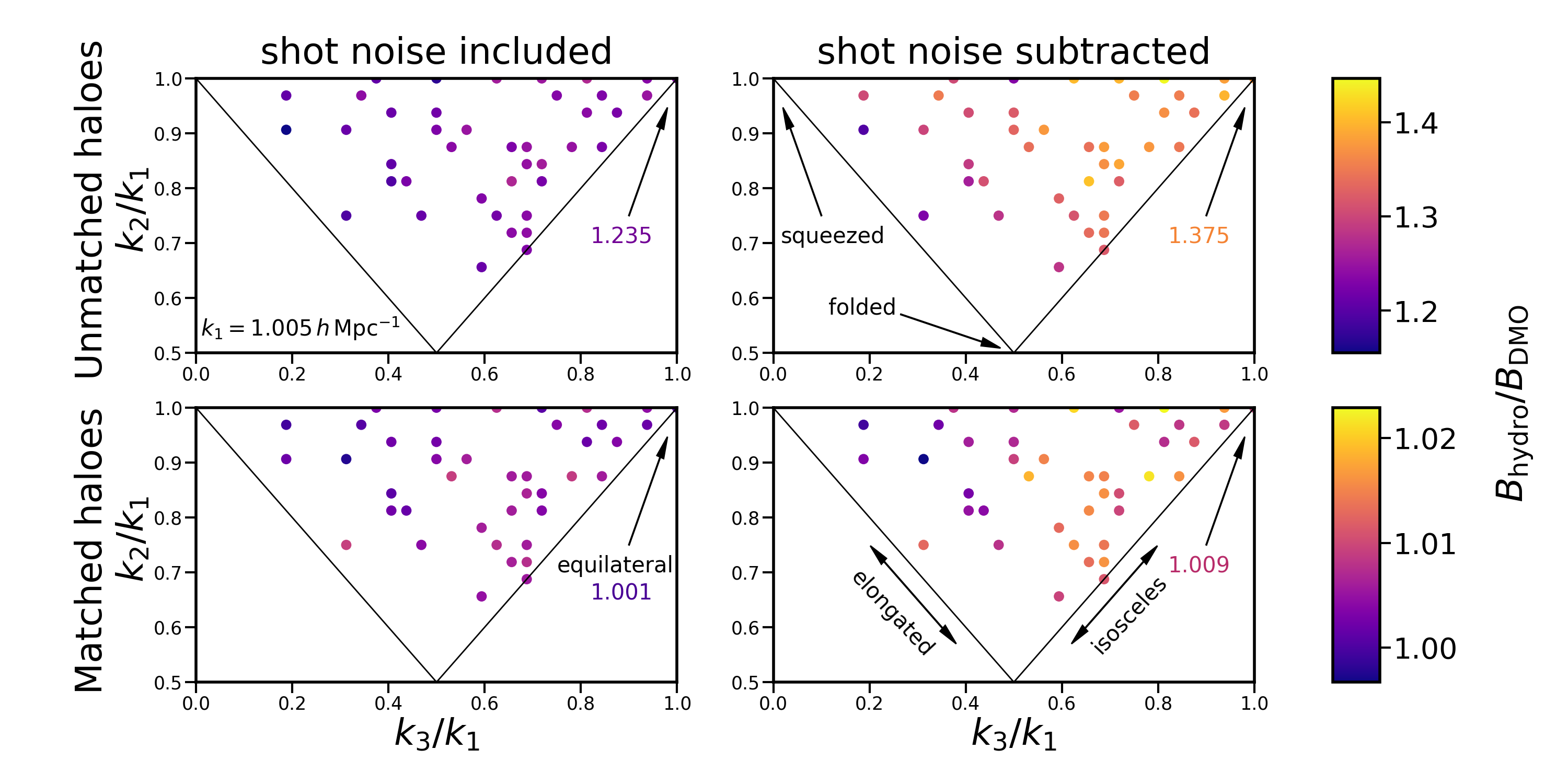}
    \caption{The same as in Fig.~\ref{fig:diff_tri_config} but for a lower value of $k_1=1.01\, \Mpch$. 
}
    \label{fig:diff_tri_config_low_k}
\end{figure*}

\begin{figure*}
	\includegraphics[width=14cm]{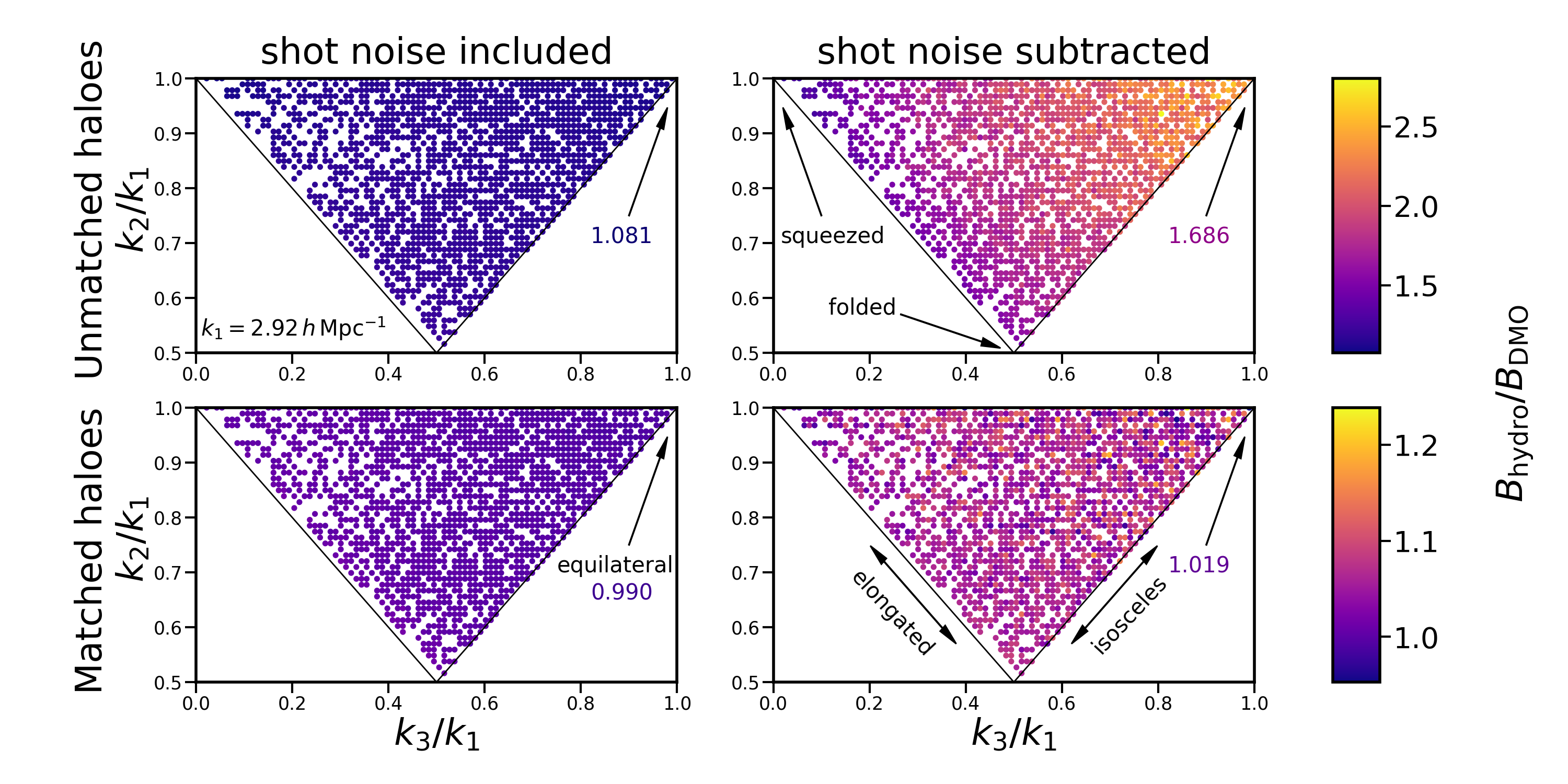}
    \caption{The same as in Fig.~\ref{fig:diff_tri_config} and Fig.~\ref{fig:diff_tri_config_low_k} but for a higher value of $k_1=2.92\, \Mpch$.
}
    \label{fig:diff_tri_config_high_k}
\end{figure*}

Here we extend the analysis presented in Section \ref{sec:Choice of triangular configuration} for different values of $k_1$.  In Section \ref{sec:Choice of triangular configuration} we adopted $k_1=1.54 \, \Mpch$, which is approximately half way between between the minimum and maximum scales we can sample in the simulation box.  Here we study the dependence of the bispectrum ratio for different triangular configurations on large scales ($k_1=1.01 \, \Mpch$; see Fig.~\ref{fig:diff_tri_config_low_k}) and for small scales ($k_1=2.92 \, \Mpch$; see Fig.~\ref{fig:diff_tri_config_high_k}).  Again the values of $k_1$ are chosen in a way that the equilateral configuration $k_1=k_2=k_3$ must exist.  For small values of $k_1$, the number of possible triangles in the box is low.  We therefore plot in Fig.~\ref{fig:diff_tri_config_low_k} the first wave vector value where the number of triangles is representative.

Overall, we see similar trends to that presented in Fig.~\ref{fig:diff_tri_config}, indicating that the dependence on the adopted triangular configuration is similar for different choices of $k_1$.


\end{document}